\documentstyle[11pt,aaspp4]{article}

\def\msun{M$_{\sun}$}
\def\uv{{\it u,v} }
\def\lsun{L$_{\sun}$}

\lefthead{Looney }
\righthead{Sub-Arcsecond Imaging of YSOs}
\begin{document}

\title{Unveiling the Circumstellar Envelope and Disk: \\
A Sub-Arcsecond  Survey of Circumstellar Structures}
\author{Leslie W. Looney\altaffilmark{1}, Lee G. Mundy}
\affil{Department of Astronomy, University of Maryland, College Park}

\and

\author{W. J. Welch}
\affil{Radio Astronomy Laboratory, University of California, Berkeley}

\altaffiltext{1}{Current Address: Max-Planck-Institut f\"{u}r Extraterrestrische
Physik, Garching, Germany\\ \hspace*{0.470cm} Email: lwl@mpe.mpg.de}

\begin{abstract}

We present the results of a $\lambda$~=~2.7~mm continuum interferometric 
survey of 24 young stellar objects in 11 fields.
The target objects range from deeply embedded Class 0 sources to optical 
T Tauri sources.  
This is the first sub-arcsecond survey of the $\lambda$~=~2.7~mm dust continuum 
emission from young, embedded stellar systems.
These multi-array observations, utilizing the high dynamic \uv range of the BIMA 
array, fully sample spatial scales ranging from 0$\farcs$4 to 60$\arcsec$,
allowing the first consistent comparison of dust emission structures in a 
variety of systems. 
The images show a diversity of structure and complexity.
The optically visible T Tauri stars (DG Tauri, HL Tauri, GG Tauri,
and GM Aurigae) have continuum emission dominated by
compact ($\le~1\arcsec$)
circumstellar disks.
In the cases of HL Tauri and DG Tauri, the disks are resolved.
The more embedded near-infrared sources (SVS13 and L1551 IRS5) have
continuum emission that is extended and compact.
The embedded sources (L1448 IRS3, NGC1333 IRAS2, NGC1333 IRAS4, VLA
1623, and IRAS 16293-2422) have continuum emission dominated by the
extended envelope, typically $\ge$ 85\% of the 
emission at $\lambda$~=~2.7~mm.
In fact, in many of the deeply embedded systems it is difficult to 
uniquely isolate the disk emission component from the envelope extending
inward to AU size scales.
Simple estimates of the circumstellar mass in the optical/infrared and
embedded systems range from 0.01-0.08 \msun and 0.04-2.88 \msun, respectively.
All of the target embedded objects are in multiple systems with separations
on scales of $\sim 30\arcsec$ or less.
Based on the system separation, we place the objects into three categories:
separate envelope (separation $\ge$~6500~AU),
common envelope (separation 150-3000 AU), 
and common disk (separation $\le$ 100 AU).
These three groups can be linked with fragmentation events 
during the star formation process:
separate envelopes from prompt initial fragmentation and the separate 
collapse of a loosely condensed cloud, common envelopes from fragmentation of a
moderately centrally condensed spherical system,
and common disk from fragmentation of a high angular momentum circumstellar
disk.

\end{abstract}


\section{Introduction}

Young stellar systems exhibit excess infrared and millimeter emission 
that arises primarily from circumstellar
dust in two basic evolutionary structures: envelopes and disks.
Current observations and theories of star formation 
(e.g. \markcite{lar69}Larson 1969; 
\markcite{pen69}Penston 1969;
\markcite{shu77}Shu 1977; 
\markcite{cas}Cassen \& Moosman 1981; 
\markcite{l84}Lada \& Wilking 1984; 
\markcite{a87}Adams, Lada, \& Shu 1987; 
\markcite{shu93}Shu et al. 1993; 
\markcite{an93}Andr\'{e}, Ward-Thompson, \& Barsony 1993)
outline an evolutionary sequence that begins with a 
density enhancement which quasi-statically contracts to form a
centrally concentrated core.
The core then gravitationally collapses forming an infall region
which feeds a central protostar and a surrounding disk supported by
centrifugal forces.  
Initially, the envelope (radii of 1000s of AU) contains most of the mass, but as the 
system evolves, the disk and protostar grow, and the disk 
becomes a significant mass reservoir.
In time, the prenatal envelope is depleted of material and progressively 
blown away by the powerful stellar outflow, revealing the young star 
and disk (radii of $\sim$~100 AU) system.
Detailed modeling of young stellar objects has shown that the 
infrared through millimeter wavelength spectral energy distribution (SED)
can be reproduced by thermal dust emission from a combination of a 
large-scale envelope, spatially thin disk, and central star 
(\markcite{a87}Adams, Lada, \& Shu 1987; 
\markcite{ken}Kenyon \& Hartmann 1987; 
\markcite{bert}Bertout, Basri, \& Bouvier 1988; 
\markcite{cal}Calvet, Hartmann, \& Whitney 1994) 
or, in some cases, by just a disk and central star 
(\markcite{b90}Beckwith et al. 1990; 
\markcite{ost}Osterloh \& Beckwith 1995; 
\markcite{d96}Dutrey et al. 1996).

Surveys of main-sequence stellar systems find that most stars
are in binary or multiple systems (\markcite{duq}Duquennoy \& Mayor 1991) with
the separation distribution ranging from a few $R_{\sun}$ to 10$^{4}$ AU
and the probability distribution peaking around 30 AU.
Recent surveys of the nearby star-forming regions of Taurus and Ophiuchus 
find that the occurrence of binaries in the young visible systems is 
about twice as common as among main-sequence stars
(\markcite{sim92}Simon et al. 1992; 
\markcite{gh93}Ghez, Neugebauer, \&, Matthews 1993; 
\markcite{le93}Leinert et al. 1993;
\markcite{rei}Reipurth \& Zinnecker 1993; 
\markcite{gh97}Ghez, White, \& Simon 1997).
How is the above star formation process altered to form binary systems, and
how does the younger, deeply embedded systems binary occurrence compare to
the young visible systems?

In this paper, we present a $\lambda$~=~2.7~mm continuum survey
of 24 nearby Young Stellar Objects that represent a sample of young 
stellar systems at various stages of evolution.
The survey highlights the large dynamic range of
\uv spacings available with the 
Berkeley-Illinois-Maryland Association (BIMA) millimeter array---
covering both the largest and smallest \uv spacings
currently available at $\lambda$~=~2.7~mm.
With a combination of low and high resolutions, we are able
to map the envelopes of the embedded sources and 
resolve-out the large-scale structure in order to peer inside 
the envelopes and image the central regions.
With the highest angular resolution to date at this wavelength,
we are able to image circumstellar disks (e.g. \markcite{mind96}Mundy et al. 1996)
and search for close binaries (e.g. \markcite{me}Looney, Mundy, \& Welch 1997).
The purpose of this paper is to present our images with the
discussion focusing on differences and similarities between the various
evolutionary stages and several broad trends in the data.
In additional papers, we will discuss individual sources in detail
and extensively model the emission structures of these sources 
as arising from extended envelopes, circumstellar disks, and circumbinary disks.

\section{Sample, Observations, and Mapping}

\subsection{Sample}

The goal of the survey was to image a range of young 
stellar systems at sub-arcsecond resolution.
We concentrated on bright, nearby sources which were
most likely to produce high dynamic range images.
The criteria for selecting the sample were:
(1) for the best sensitivity to solar-system-sized spatial scales,
we focused on the closest sources ($\le$ 350 pc);
(2) to insure sufficient surface brightness at 
sub-arcsecond resolution, we chose among the brightest sources 
in the nearby star forming regions;
(3) to survey a number of evolutionary stages,
our sample included the youngest, most deeply embedded 
sources (Class 0), as well as optical T Tauri stars (Class I/II);
(4) finally, to achieve the best map fidelity,
we required that the sources have both a 
bright phase reference calibrator and a weaker point source 
response calibrator nearby on the sky.
This combination of criteria introduces three strong biases in our sample:
(a) we favor bright millimeter sources which should be systems with more
circumstellar material than average (e.g. \markcite{b90}Beckwith et al. 1990),
(b) most of our sources are either Class 0 or Class II (this is mostly due to
the bias above), and 
(c) as a result of our distance limit, our sources are drawn from
just three local clouds--- Taurus, Ophiuchus, and Perseus.
Table 1 lists the sources, adopted distances,
whether they are optically visible (in this 
category we also include objects which are visible in the near-infrared) 
or embedded, adopted SED Class, main calibrator, secondary 
calibrator, and distance reference.
The distance to the Perseus objects has been in dispute lately,
350 pc (\markcite{herb}Herbig \& Jones 1983) and 
more recently 220 pc (\markcite{cer}Cernis 1990).  
We adopt the previous value, allowing an easier comparison
to earlier works.

The frequency of multiple systems will be addressed in \S 6, but 
the number of objects in the embedded source group
is misleading since all of the observed embedded sources are in 
multiple systems.
For example, the Perseus objects include 13 sources which were
mapped in only 4 pointings,
whereas in the optical group, only  DG Tauri and DG Tauri B
(which is an embedded object not thought to be related to DG Tauri
except by projected proximity;
\markcite{jone}Jones \& Cohen 1986) were mapped in the 
same observation.

\subsection{Observations}

All sources were observed in three configurations (C, B, and A) of the
9-element BIMA Array\footnotemark (\markcite{welch}Welch et al. 1996).
The observations were acquired from 1996 May to 1998 March.
The digital correlator was configured with two 700~MHz bands
centered at 106.04~GHz and 109.45~GHz.
The C$^{18}$O(1-0) line was observed simultaneously;
those results will be discussed elsewhere.
The two continuum bands were checked for consistency, then
combined in the final images.
The system temperatures during the observations ranged from
150-700 K (SSB). 

\footnotetext{ The BIMA Array is operated by the Berkeley Illinois Maryland
Association under funding from the National Science Foundation.}

In the compact C array (typical synthesized beam of
$\sim$8$\arcsec$), the shortest baselines were limited
by the antenna size of 6.1~m, yielding
a minimum projected baseline of 2.1~k$\lambda$ and good
sensitivity to structures as large as $\sim$50$\arcsec$.
In the mid-sized B array (typical synthesized beam of
$\sim$2$\arcsec$), the observations are sensitive to
structures as large as $\sim$10$\arcsec$.
In the long baseline A array (typical synthesized beam of
$\sim$0$\farcs$6), the longest baselines were typically 
450~k$\lambda$.
The combination of these three arrays provides a 
well sampled \uv plane from 2.1~k$\lambda$ out to 400~k$\lambda$.

The uncertainty in the amplitude calibration is 
estimated to be 20\%.
In the B and C arrays, the amplitude calibration was boot-strapped
from Mars.  In the A Array, amplitude calibration was done
by assuming the flux of the quasar 3C273 to be 18.8 Jy at the end of
1996 October, 23.0  Jy at the end of 1996 November, 
and 27.0 Jy until the end of 1997 January.
Absolute positions in our map have uncertainty due to the
uncertainty in the antenna locations and the statistical variation
from the signal-to-noise of the observation. These two
factors add in quadrature to give a typical absolute
positional uncertainty of $0\farcs$15 in the highest resolution maps.

The A array observations required careful phase calibration.
On long baselines, the interferometer phase is very sensitive
to atmospheric fluctuations.  
We employed rapid phase referencing; the observations were switched 
between source and phase calibrator on a two minute cycle, to follow 
the atmospheric phase 
(\markcite{hold}Holdaway \& Owen 1995; \markcite{me}Looney, Mundy, \& Welch 1997).
To monitor the coherence of the atmosphere, or the ``seeing'',
we included a nearby weaker quasar in the 
observation cycle.
This quasar was imaged along with the target source
as a check of the point source response
and for accurate image registration.
In the observations presented here, the secondary quasar always
mapped as a point source within statistical uncertainties.

\subsection{Displaying the Data}

The observational data span \uv distances from 2.1k$\lambda$
to 450k$\lambda$, providing information of
the brightness distribution on spatial scales from
0$\farcs$4 to 60$\arcsec$.
In order to display this information in the image plane, we 
have mapped the emission with four different \uv
weighting schemes which stress structures
on spatial scales of roughly 5$\arcsec$, 3$\arcsec$,
1$\arcsec$, and 0$\farcs6$. 
These resolutions were typically obtained 
with natural weighting, robust weighting (\markcite{brig}Briggs 1995) of 1.0, 
robust weighting of 0.0, and robust weighting of -0.5, respectively.

\section{Results}

The $\lambda$~=~2.7~mm continuum images from the survey
are shown in Figures 1 through 16, first the optical/IR systems
and then the deeply embedded systems.
In each figure, the four panels display the same multi-configuration data
with different \uv weighting schemes to emphasize structures
on size scales comparable to the synthesized beam.
Table 2 lists, at each resolution, the peak flux, the integrated flux, and the
box used to measure the integrated flux.
The error bars on the flux measurements represent the statistical uncertainties.

In Figure 17, the source fluxes are presented in a plot comparing
the total integrated flux to the ratio of the visibility amplitude 
at two specific fringe spacings.
The horizontal axis is the integrated flux of each object taken from
Table 2 and adjusted to the distance of Taurus (140 pc).
For the vertical axis, the \uv data were binned in annuli stretching from
2.2 k$\lambda$ to 7.8 k$\lambda$ and from 22 k$\lambda$ to 78 k$\lambda$,
corresponding to average spacings of 5 k$\lambda$ and 50 k$\lambda$ for 
the distance of Taurus.
The ranges were chosen to provide the best signal to noise for a
typical source and the best sensitivity to large and small scale 
structures.
The vertical axis plots the ratio of the vector averaged amplitudes in the 
two bins or the 5k$\lambda$/50k$\lambda$ ratio.
In order for the ratio to consistently probe the same spatial scales for 
all objects,
we adjusted the bin range to take into account the source distance:
for $\rho$ Ophiuchi we used annuli averaging of 5.7 k$\lambda$ 
and 57 k$\lambda$, respectively,
and for Perseus we used annuli averaging 
12.5 k$\lambda$ and 125 k$\lambda$, respectively.
For objects in multiple systems, the companion objects were subtracted out 
of the \uv data using the CLEAN components in the lower
resolution (5$\arcsec$ and 3$\arcsec$) maps as a model.

Figure 18 illustrates how to interpret Figure 17 by comparing
the visibility curves of a 5000 AU
outer radius envelope and a 100 AU radius disk, both at the distance of
Taurus (140 pc).
This figure is essentially the one-dimensional Fourier Transform of the 
brightness distribution of the source on the sky.
The horizontal axis is the spatial frequency of the interferometer, or the
antenna separation in units of wavelengths, and the vertical axis
is in Janskys.
The visibility of the disk model is similar to an unresolved point source 
until the disk starts to be resolved at $\sim$ 60 k$\lambda$.
In the case of the envelope, it is quickly resolved and approaches the
expected power-law form; the ``wiggles'' are due to the sharp outer edge
of the model.
Overall, the 5k$\lambda$/50k$\lambda$ ratio provides a measure of the 
relative emission on spatial scales of
2500 - 9000 AU and 250 - 900 AU and quantitatively measures the
``compactness'' of the object.
An object with spatially extended structure, such as an envelope with
size scales of 1000 AU or larger will have a 5k$\lambda$/50k$\lambda$ 
ratio $>$1, and
an object that is entirely compact, such as a circumstellar
disk with radius of $\sim$ 100 AU, will be essentially unresolved
in the two bins and the ratio will be unity.

The general trend of Figure 17 is, as expected, that most of the optical sources
(solid triangles) are 
compact (5k$\lambda$/50k$\lambda$ ratio of $\sim$ 1) and most of the 
embedded sources
(solid squares) are surrounded by envelopes that are being 
significantly resolved at 50 k$\lambda$ (5k$\lambda$/50k$\lambda$ ratio $>$1).
However, there are a couple of exceptions worth discussing.
First, there are two optical stars with unusually 
large 5k$\lambda$/50k$\lambda$ ratios--- GG Tauri and
SVS13 A.
GG Tauri is a close binary system with a separation of 0$\farcs$255
(\markcite{le91}Leinert et al. 1991) and a large circumbinary disk 
(diameter $\sim$ 400 AU;
\markcite{sim92b}Simon \& Guilloteau 1992; 
\markcite{d94}Dutrey, Guilloteau, \& Simon 1994).
Thus, in GG Tauri the 5k$\lambda$/50k$\lambda$ ratio is resolving the large scale circumbinary disk.
SVS13 A, first detected in the infrared at 2.2 $\mu$m 
(\markcite{str74}Strom, Grasdalen, \& Strom 1974; 
\markcite{str76}Strom, Vrba, \& Strom 1976) is also know to have 
optical/infrared outbursts (\markcite{eis}Eisl\"{o}ffel 1991); 
yet it has a very large 5k$\lambda$/50k$\lambda$ ratio.
There are two possible explanations: (1) the envelope of the nearby, younger embedded 
object SVS13 A2 is contributing to the flux of SVS13 A1 at 5 k$\lambda$,
or (2) the optical/infrared emission is
a reflection nebula and the source should be classified as embedded.
A second set of exceptions are two embedded sources 
(NGC 1333 IRAS2-B and IRAS4-C) that have unusually 
small 5k$\lambda$/50k$\lambda$ ratios.  
Since these sources are not detected in the optical or the near-infrared
they are classified as embedded sources, but 
their 5k$\lambda$/50k$\lambda$ ratio and their 
integrated fluxes in Table 2 indicate that they are compact.
These two sources could be optical/near-infrared sources that 
are viewed through intervening obscuration.
The following subsections discuss each of the sources in more detail.

\subsection{DG Tauri}

DG Tauri is a well studied classical T Tauri star system.
Through modeling of the system's SED, 
\markcite{a90}Adams, Emerson, \& Fuller (1990) estimated a
radius of 75 AU for the circumstellar disk.
The source was observed in the near-infrared during a lunar 
occultation (\markcite{le91}Leinert et al. 1991), 
and it was determined that the star was a single 
system with an extended ``shell'' 6.8 AU in diameter.
In addition, near-infrared speckle observations revealed the presence of a 
``halo'' with a diameter of 130 AU (\markcite{le91}Leinert et al. 1991).
In panel (d) of Figure 1, the circumstellar disk around DG Tauri
is marginally resolved.
Fitting an elliptical Gaussian to the image in panel (d),
we obtain a deconvolved Gaussian size of 0$\farcs$61 $\pm$ 0$\farcs$1
$\times$ 0$\farcs$57 $\pm$ 0$\farcs$1 with a principal axis of
167$\arcdeg$ $\pm$ 10$\arcdeg$.
This result is different from the measurement at $\lambda$~=~2.0~mm
from \markcite{kit}Kitamura, Kawabe, \& Saito (1996), which 
found a deconvolved size of 1$\farcs$56 $\times$ 0$\farcs$54 at a principal 
axis of 99$\arcdeg$, but similar to the measurement at $\lambda$ = 2.7 mm
from \markcite{d96}Dutrey et al. (1996), which found a deconvolved size of 1$\farcs$1 
$\times$ 0$\farcs$6 at a principal axis of  150$\arcdeg$.
The extension to the southwest in panel (d) lies along the direction of 
the jet (\markcite{cb86}Cohen \& Bieging 1986; 
\markcite{sta97}Stapelfeldt et al. 1997; 
\markcite{la}Lavalley et al. 1997) and may trace ionized gas in the jet.

\subsection{DG Tauri B} 

DG Tauri B, located 53$\arcsec$ southwest of DG Tauri,
has a molecular outflow (principal axis of $\sim$ 295$\arcdeg$;
\markcite{mi94}Mitchell et al. 1994; 
\markcite{mi97}Mitchell, Sargent, \& Mannings 1997) that is driven 
by a jet seen at optical (\markcite{mun87}Mundt, Brugel, \& B\"{u}hrke 1987) 
and centimeter
(\markcite{rod95}Rodr\'{i}guez, Anglada, \& Raga 1995) wavelengths.
DG Tauri B was observed near the half power point of our beam during 
the observation of DG Tauri; therefore 
measured fluxes have a significant additional uncertainty.
The fluxes listed in Table 2 were corrected for the primary beam attenuation.

The morphology of DG Tauri B changes in Figure 2 with increasing
resolution.
Going from panel (d) to panel (c) to panel (b), the major elongation of the
emission changes from northwest to north to slightly northeast.
In panel (a) the emission is triangular with extension to the northwest, northeast,
and southwest.
The simplest interpretation is that the high resolution image traces the 
ionized gas, while the lower resolution images trace both ionized gas and dust.
The position angle for the larger scale dust emission is then $\sim$35$\arcdeg$,
which is consistent with the optical extinction lane 
(\markcite{sta97}Stapelfeldt et al. 1997) and perpendicular to the outflow jet.
The extended emission in panel (d) strongly resembles the $\lambda$ = 3.6 cm
image (\markcite{rod95}Rodr\'{i}guez, Anglada, \& Raga 1995), suggesting 
that it is tracing ionized gas in the jet.
The relative flux numbers in Table 2 suggest that roughly half
of the flux arises from dust and half from ionized gas in the jet.

\subsection{L1551 IRS5}

L1551 IRS5, one of the prototypical Class I sources in the classification scheme
of \markcite{a87}Adams, Lada, \& Shu (1987), has the most 
spectacular bipolar molecular outflow 
in the Taurus cloud (principal axis of $\sim$ 50$\arcdeg$; 
\markcite{snell}Snell, Loren, \& Plambeck 1980).
The $\lambda$ = 2.7 mm continuum data presented here were discussed in detail by
\markcite{me}Looney, Mundy, \& Welch (1997).
The source is argued to be a proto-binary system with a large-scale 
envelope ($\sim$1300 AU radius), circumbinary disk, and two circumstellar disks
(separation of 0$\farcs$35).
The binary circumstellar disks have recently been resolved in 
$\lambda$~=~7~mm observations (\markcite{rod98}Rodr\'{i}guez et al. 1998).
In Figure 3 panels (a) and (b), the emission is dominated by the large-scale envelope, while
panel (c) clearly shows the circumbinary envelope.
In panel (d), the two point-source-like circumstellar disks are still
convolved with the low-level emission from the circumbinary envelope
which is extended along a principal axis of $\sim$~160$\arcdeg$.
The higher resolution image from 
\markcite{me}Looney, Mundy, \& Welch (1997) is not
shown.

\subsection{HL Tauri}

HL Tauri, perhaps the most studied of the optical/IR visible young stars,
has a large-scale CO structure ($\sim$~1500 AU, 
\markcite{sar}Sargent \& Beckwith 1991; \markcite{hay}Hayashi, Ohashi, \& Miyama 1994) 
and a compact circumstellar disk ($\sim$ 100 AU) that has been
resolved by the CSO-JCMT interferometer (\markcite{lay94}Lay et al. 1994; 
\markcite{lay97}Lay et al. 1997) 
and imaged by the BIMA array (\markcite{mund96}Mundy et al. 1996).
Figure 4 shows the new BIMA image which has both lower noise and
higher resolution than the image of \markcite{mund96}Mundy et al. (1996).

In panel (d), the circumstellar disk of HL Tauri is clearly evident.
Fitting an elliptical Gaussian to the image, we obtain a deconvolved 
Gaussian size of 0$\farcs$88$\pm$0$\farcs$1 $\times$ 0$\farcs$58$\pm$0$\farcs$1 
and principal axis of 111$\arcdeg$$\pm$10$\arcdeg$, which agrees with the 
observations of \markcite{lay94}Lay et al. (1994) and 
\markcite{mund96}Mundy et al. (1996).
However, fitting an elliptical Gaussian to the image is not 
the correct method for determining the true disk size.
Recent modeling of the HL Tauri circumstellar disk found that simple
models could not fit the CSO-JCMT single baseline interferometer
$\lambda$ = 650 $\mu$m and 850 $\mu$m data and the 
$\lambda$ = 2.7 mm and 7 mm data (\markcite{lay97}Lay et al. 1997).
We will consider more complicated disk models in a subsequent paper.
The image in panel (d) also shows an extension to the north-east along
the axis of the optical jet, principal axis 46$\arcdeg$ 
(\markcite{mun90}Mundt et al. 1990).
This feature likely arises from free-free emission in the jet; such
free-free emission dominates the highest resolution maps at
$\lambda$ = 7 mm (\markcite{wil99}Wilner et al. 1999).

HL Tauri is classified as an optical source, but has recently been
shown to be embedded within a reflection nebula 
(\markcite{sta95}Stapelfeldt et al. 1995);
we do not see the star directly in optical light,
but it can be seen directly in the near-infrared 
(\markcite{wein}Weintraub, Kastner, \& Whitney 1995;
\markcite{b95}Beckwith \& Birk 1995).
Our data do not conclusively detect envelope emission
associated with the extended nebula.
The envelope on size scales larger than 3$\arcsec$ contributes less than 10$\%$ of 
the dust emission, where our estimate is limited by the uncertainty
of the relative amplitude calibration between arrays .

\subsection{GG Tauri}

GG Tauri is a close binary system with a separation of 0$\farcs$25
(\markcite{le91}Leinert et al. 1991) and a large circumbinary ``ring-like'' disk 
(diameter $\sim$ 400 AU;
\markcite{sim92b}Simon \& Guilloteau 1992; 
\markcite{d94}Dutrey, Guilloteau, \& Simon 1994).
Our images presented in Figure 5, have different \uv weighting schemes
from the rest of the surveyed objects stressing size scales of
5$\arcsec$, 2$\arcsec$, 1$\arcsec$, and 0$\farcs$9.
Fitting an elliptical Gaussian to the image in panel (b), we obtain
a deconvolved size of 3$\farcs$3 $\pm$ 0$\farcs$1 $\times$
2$\farcs$7 $\pm$ 0$\farcs$1 
at a position angle of 82$\arcdeg$ $\pm$ 10$\arcdeg$,
which is in good agreement with 
\markcite{d94}Dutrey, Guilloteau, \& Simon (1994).

Most of the peaks and valleys in the emission in
panel (d) represent $\le$2.5$\sigma$ variations from an overall
flat structure; hence, it is difficult to say whether they are
real variations in the surface density or temperature.
The emission is not strongly concentrated toward the inner
edge of the ring indicating that there is not a strong surface density
gradient within the disk. A more detailed discussion of the disk will be 
presented in \markcite{mund99}Mundy, Looney, \& Welch (1999).
No emission is detected from the system at 0$\farcs$6 resolution;
this places upper limits on the emission from any
compact structures ($<$ 0$\farcs$6), such as individual circumstellar disks 
within the binary system, at a 3$\sigma$ limit of 5 mJy.
The companion binary system of this quadruple system, GG Tauri/c,
was not detected at any resolution; the 3$\sigma$ upper limit on its 
flux density is 4 mJy.

\subsection{GM Aurigae}

GM Aurigae is another classical T Tauri star system that has a large-scale
CO structure (\markcite{k93}Koerner, Sargent, \& Beckwith 1993).
In Figure 6 panel (d), we do not see evidence that the disk is resolved, but the 
signal-to-noise is poor.
Fitting an elliptical Gaussian to the image in panel (d) yields
a point source.
We can place a limit on the deconvolved Gaussian size for the circumstellar 
disk of $\le$0$\farcs$4, at the 95\% confidence level.
In panel (c), the emission seems slightly extended along the direction 
perpendicular to the larger scale CO structure which has a position angle
of 55$\arcdeg$.

The total flux density reported in Table 2 (22 mJy) is roughly consistent
with that measured by \markcite{d96}Dutrey et al. (1996), 28 mJy.
Unlike \markcite{d96}Dutrey et al., we do not directly resolve the disk.
However, we do measure a 35\% decrease in flux density between the 5$\arcsec$
and 0$\farcs$6 beams, indicating that some structure is present.
 
\subsection{L1448 IRS3 Region}

The L1448 complex is located about $\sim$1$\arcdeg$ southwest of NGC 1333.
IRAS revealed three strong infrared sources in the region, of which L1448 IRS3 
was the brightest in the far infrared 
(\markcite{ba86}Bachiller \& Cernicharo 1986).
IRS3 is projected within the blueshifted lobe of the impressive, highly 
collimated molecular outflow from L1448-mm which lies to the southeast
(\markcite{ba90}Bachiller et al. 1990; 
\markcite{ba91}Bachiller, Andr{\`e}, \& Cabrit 1991).
Coinciding within the uncertainties of the L1448 IRS3 source is
a very strong H$_{2}$O maser and $\lambda$ = 6 cm compact emission
(\markcite{ang}Anglada et al. 1989).
Higher resolution maps in the $\lambda$ = 2 cm and 
6 cm continuum found that the source was composed of two sources
L1448 N(A) and L1448 N(B) (\markcite{cur}Curiel et al. 1990;
\markcite{bar} Barsony et al. 1998).
Curiel et al. separated the region into two areas: L1448 C, the center of the
molecular outflow, and L1448 N corresponding to the IRS3 source.
L1448 N(B) contributes most of the flux at millimeter wavelengths 
(\markcite{ter93}Terebey, Chandler, \& Andr\'{e} 1993; 
\markcite{ter97}Terebey \& Padgett 1997).
A third source is also seen at $\lambda$ = 2.7 mm which lies to the north-west
of L1448 N(B) (\markcite{ter97}Terebey \& Padgett 1997).

In our images of the region, we clearly detect all three sources
which we label:
L1448 IRS3 A, B, and C, using the IAU nomenclature.
The three sources are indicated in Figure 7 (b).
Source A, which is the brightest source at centimeter wavelengths,
is significantly weaker than source B at $\lambda$ = 2.7 mm.
In fact at the highest resolution, source A is not detected.
However, the spectral index derived by \markcite{cur}Curiel et al.
for source A ($\alpha \sim$ 0.2) suggests only $\sim$ 2 mJy of the
emission arises from free-free emission. Thus, the millimeter wavelength
flux from IRS3~A is likely dominated by dust emission.

Located to the north-west, source C is detected at all resolutions.
Unfortunately, source C is too weak to be plotted
on the 5k$\lambda$/50k$\lambda$ ratio figure.
In panels (c) and (d), source B shows very complicated morphology 
on small scales.
There is an outflow associated with the IRS3 region which is 
nearly parallel to the outflow from L1448-mm, at a position angle
of $\sim$ -21$\arcdeg$ (\markcite{ba95}Bachiller et al. 1995; 
\markcite{dav}Davis \& Smith 1995).
The extension that is seen in panel (c) and (d) is almost
perpendicular with the outflow, but it is unclear if it is an
envelope or a large disk. 
The peak flux density decreases by a factor of two in each step of resolution in
Figure 7(b), to (c), to (d), so there is considerable
structure on all spatial scales.

\subsection{NGC 1333 IRAS2}

The NGC 1333 star forming region in Perseus is an extremely active site
of star formation with multiple infrared sources 
(\markcite{str76}Strom, Vrba, Strom 1976;
\markcite{asp}Aspin, Sandell, \& Russell 1994; 
\markcite{l96}Lada, Alves, \& Lada 1996) and outflows
(\markcite{san94}Sandell et al. 1994; 
\markcite{war}Warin et al. 1996; \markcite{ball}Bally et al. 1996).
NGC 1333 IRAS2 (\markcite{jenn}Jennings et al. 1987) is located
on the edge of the large cavity in NGC1333 
(\markcite{lan}Langer, Castets, \& Lefloch 1996).
The region has a two outflows that originate near IRAS2:
the ``N-S'' outflow with principal axis of $\sim$ 25$\arcdeg$
(\markcite{lis}Liseau, Sandell, \& Knee 1988) and the 
``E-W'' outflow with principal axis
of $\sim$ 104$\arcdeg$ (\markcite{san94}Sandell et al. 1994).
Recent millimeter interferometric observations show that there are two 
continuum peaks that are probably associated with the two outflows,
and that the northern source (Source A) is responsible for the ``E-W''
outflow (\markcite{bla}Blake 1997).

Figures 8 and 9 show NGC 1333 IRAS2 A and B respectively.  
Source A, the stronger of the two sources, is mostly extended emission,
but the flux remaining in panel (d) is consistent with a point source.
Source B is mostly compact emission.
The extension of source B in panel (d) is nearly perpendicular with the
``N-S'' outflow, suggesting a possible circumstellar structure.

\subsection{SVS 13}

Located northwest of IRAS2, the young stellar object SVS13 
(\markcite{str76}Strom, Vrba, Strom 1976; also referred to as 
SSV13 in the literature from \markcite{herb}Herbig \& Jones 1983)
is associated with the NGC 1333 IRAS3 region (\markcite{jenn}Jennings et al. 1987).
IRAS3 is comprised of at least 3 millimeter sources:
source A located near the infrared position of SVS13, source B to the southwest
(\markcite{gross}Grossman et al. 1987; 
\markcite{chini}Chini et al. 1997;
\markcite{ba98}Bachiller et al. 1998), and source C further to the southwest
(\markcite{chini}Chini et al. 1997; \markcite{ba98}Bachiller et al. 1998).

Figures 10 and 11 clearly show all three millimeter sources.
In panel (b) of both figures there is another source located to the 
southwest of source A.
This source (which we will call A2) is coincident with VLA source 3
from recent VLA observations of this region 
(\markcite{rod97}Rodr\'{i}guez, Anglada, \& Curiel 1997).
Located $\sim$ 6$\arcsec$ from SVS13, Rodr\'{i}guez et al. argue that
A2 is a better candidate for the HH 7-11 outflow 
(\markcite{rod97}Rodr\'{i}guez, Anglada, \& Curiel 1997).
However, source A2 is only a 3$\sigma$ detection in panel (c) and is 
not detected at higher resolution.
We suggest that its lack of compact structure makes it a less likely 
candidate for driving the outflow, despite the fact that the centimeter
emission suggests that it also has a jet.
In addition, the high spectral index in the centimeter ($\alpha \sim$ 1.5;
\markcite{rod97}Rodr\'{i}guez, Anglada, \& Curiel 1997)
and the lack of detection in high resolution $\lambda$~=~1.3~mm observations
(\markcite{ba98}Bachiller et al. 1998)
suggest that this source may be dominated by free-free emission.
Source A1 is coincident with the infrared/optical source SVS13.
Since source A1 is an optical source, we would expect it to be an older object.
However, our data suggest that A1 is deeply embedded in an envelope.
The SVS13 results are discussed in detail in 
\markcite{welch99}Welch, Looney, \& Mundy (1999).

\subsection{NGC 1333 IRAS4}

One of the most well known sources in the NGC 1333 region is the object 
NGC 1333 IRAS4, located to the southwest of SVS13.
Unresolved in the IRAS images (\markcite{jenn}Jennings et al. 1987), IRAS4 
breaks into two bright objects at sub-millimeter wavelengths 
(\markcite{san91}Sandell et al. 1991).
Our images, Figures 12, 13, and 14, show three objects: IRAS4 A, B, and C.
Our data provide the first indication that source C may be a young star.
Source C is detected at all resolutions, has 
a 5k$\lambda$/50k$\lambda$ ratio near 1,
and the integrated flux in Table 2 is constant at all resolutions.
The source C characteristics are more like those of an
optical/IR source than its IRAS4 companions.

NGC1333 IRAS4 A \& B have been observed with the CSO-JCMT single baseline
interferometer at $\lambda$ = 840 $\mu$m (\markcite{lay95}Lay, Carlstrom, \& Hills 1995).
Their best fit for source A was two elliptical Gaussians, and indeed,
in our images source A is a binary system.
It is interesting to note that the best fit fluxes from Lay et al. give a
ratio of 0.78, while our two sources have a flux ratio of 0.25.
This suggests that either the dust emissivity of these two objects vary
differently with frequency or the optical depth is very different.
For source B, the CSO-JCMT data could not be fit with a single star
or binary model.  
Lay et al. suggest that source B may be a triple system; however, they were 
not aware of source C at that time, which may have confused their analysis.
Our image of source B shows weak extensions (4$\sigma$) to the north 
and southwest, but our data are not sufficient to determine the nature 
of these features.
They could trace a multiple stellar system or inhomogeneities within the
envelope.

\subsection{VLA 1623}

The source VLA 1623, near the center of the $\rho$ Ophiuchi cloud core A,
is the prototypical Class 0 source 
(\markcite{an93}Andr{\`e}, Ward-Thompson, \& Barsony 1993) 
that drives a large outflow with
principal axis $\sim$~-60$\arcdeg$ (\markcite{an90}Andr{\`e} et al. 1990; 
\markcite{den}Dent et al. 1995; 
\markcite{yu}Yu \& Chernin 1997).
This source has been observed with the CSO-JCMT single baseline interferometer 
at $\lambda$~=~1360 and 845 $\mu$m (\markcite{pud}Pudritz et al. 1996).
They modeled the source as a Gaussian and placed a 70 AU radius 
upper limit on the size of the compact circumstellar disk.

Recent, high resolution VLA observations at $\lambda$~=~3.6~cm 
(\markcite{bon}Bontemps \& Andr{\`e} 1997)
show a series of emission clumps that are interpreted as knots of
a radio jet driving the large CO outflow.
However, the position angle of the radio jet and the CO outflow differ by 
$\sim$ 30$\arcdeg$.
In our highest resolution images, Figure 15 panels (c) and (d), the 
millimeter emission breaks into nearly equal point sources.
The two crosses mark the positions of the two centimeter sources from
\markcite{bon}Bontemps \& Andr{\`e} (1997) that appear associated with 
the millimeter 
emission; the source positions at the two wavelengths agree to 
within the uncertainties.
The total emission from the two sources at $\lambda$ = 3.6 cm is 
less than 1 mJy, so free-free emission is probably not contributing
much of the 70 mJy seen at $\lambda$ = 2.7 mm.
In addition, we have reanalyzed the data of 
\markcite{pud}Pudritz et al. (1996) and find that a binary interpretation 
is also consistent with their data due to their limited \uv sampling.
Thus, the emission at $\lambda$ = 2.7 mm is probably
dominated by dust emission.
VLA 1623 is most likely a very young binary system
with two circumstellar disks.
Like IRAS 16293-2422, we refer to the southern source as A
and the northern source as B.

\subsection{IRAS 16293-2422}

IRAS 16293-2422 is a very well studied deeply embedded binary
system with two molecular outflows
(\markcite{walk}Walker et al. 1986; 
\markcite{woot}Wootten 1989; \markcite{mund92}Mundy et al. 1992;
\markcite{walk93}Walker et al. 1993) 
in $\rho$ Ophiuchi.
The outflow from the southern source A has a principal axis of 
$\sim$ 50$\arcdeg$, and the outflow of the northern source B
has a principal axis of $\sim$ 75$\arcdeg$.
The outflow from source B does not extend down near to the source, which may 
indicate that source B is no longer driving its outflow.  
In high resolution observations at $\lambda$ = 2 cm, the system is comprised
of three peaks: A1 and A2 to the southeast and B to the northwest
(\markcite{woot}Wootten 1989).
In Figure 16, we detect the two sources, A and B, that were detected 
previously at $\lambda$ = 2.7 mm (\markcite{mund86}Mundy et al. 1986; 
\markcite{mund92}Mundy et al. 1992).
In panel (c), there is still a clear connection between the two sources that is 
most likely a circumbinary envelope.
In panel (d), the massive circumbinary envelope is mostly
resolved-out and the residual emission arises from two compact sources
and some weak extensions that are probably attributed to
density enhancements within the circumbinary structure.
At high resolution, source A appears elongated along the position angle of the
$\lambda$ = 2 cm sources, which are indicated in panel (d) as crosses.
IRAS 16293-2422 source A has the highest 
5k$\lambda$/50k$\lambda$ ratio in the survey.
In fact, the ratio is twice as large as the next highest source 
L1448 IRS3 B.
Thus, source A has most of its mass in the envelope, 
perhaps making it one of the younger sources in this survey.

Our measurement of the integrated flux from this source is higher than
previous observations (e.g. \markcite{walk93}Walker et al. 1993).
This is because we have shorter spacing \uv data which pick up
the extended structure of the circumbinary envelope better than previous works.
If we remove the shorter \uv spacings, the total integrated
flux is $\sim$750 mJy, which is more in agreement with the previous
measurements.

\section{Comparison of Structure}

There is a striking difference between the embedded objects and the optical/IR
objects in our survey.
The optical sources have compact emission on spatial scales of 
$\sim$1$\arcsec$ with little large-scale envelope emission.
This is illustrated both in Figure 17 and by the peak/integrated fluxes
in Table 2. 
The peak flux does not change significantly, even down to size scales of 
$\sim$1$\farcs$5, until the resolution is sufficient to see the circumstellar disk.
This contrasts strongly with the embedded objects which typically have
$\geq  85\%$ of their emission in large scale structures.
The embedded sources, Figures 7 through 16, show dramatic change in structure 
as the resolution is varied through the panels.
Structures are resolved-out as the shorter \uv spacings
are down-weighted in the higher resolution images.
At the highest resolution, the embedded objects
typically have a residual compact component, but the flux of this component
is significantly less than that in large scale extended emission.
In addition, the embedded objects often show complex sub-structure
within the field.

Indeed, it is usually difficult to isolate the
circumstellar disk from the envelope in the embedded sources,
even at 0$\farcs$5 resolution.
In most cases, the emission shows no discontinuity in flux between
1$\arcsec$ and 3$\arcsec$ scales; this indicates that any disk
present cannot be significantly more massive than the mass of the
envelope extended to small scales. Since the embedded sources are typically
a factor of two farther away than the optical sources,
we cannot establish whether the disks in embedded systems
are systematically less massive than typical disks around
young optical stars.
Nonetheless, the younger disks do not appear to be prominent mass reservoirs
compared to their envelopes on the few hundred AU scale.

However, it is interesting to note that even in the two nearest embedded 
systems VLA 1623 and IRAS 16293-2422, where our spatial resolution is
comparable to the Taurus systems, the disk and envelope contributions to the
continuum emission are difficult to disentangle.
In the VLA 1623 system, roughly one-third of the total flux is in the 
large-scale circumbinary envelope, and two-thirds is in the 
two unresolved components ($\le$ 160 diameter);
each compact component may be a $\sim$0.04\msun\ circumstellar disk or 
a combination of an individual circumstellar envelope and a circumstellar disk. 
In the IRAS 16293-2422 system, most of the flux ($\sim$85\%) is in the large
scale circumbinary envelope and any disk flux is less than 15\% of the 
total flux, but even here the envelope emission could still be significant
fraction of the flux at small scales.

Overall, our results generally support a picture in which most embedded systems 
have disk masses which are comparable to, or less than, the
typical disk masses of $\sim$0.02~\msun\ 
found for young optical stars (\markcite{ost}Osterloh \& Beckwith 1995). 
Most theoretical works find that the disk grows in prominence with time
as the system evolves.
\markcite{cas}Cassen \& Moosman (1981) showed that the detailed evolution of a 
disk is very dependent upon the distribution of mass and angular 
momentum in the original cloud and dissipative processes within the disk.
For a range of assumptions, they found that the disk should
grow more massive and larger with time.
Building upon these results, \markcite{stah}Stahler et al. (1994) 
considered a disk
with negligible viscosity that was formed as soon as the angular momentum
in the infalling material caused it to ``miss'' the protostar.
They found that the radius of the disk is a strong function of time, increasing
as $t^{3}$.  As the accretion rate onto the star begins to fall off,
the mass of the disk increases.
Recent results using magnetic collapse models 
(\markcite{basu}Basu 1998), predict that the early disk grows less strongly,
increasing only linearly with time, but, in these models, the disk starts out
more massive since less material can fall directly onto the star.
Unfortunately, even higher linear resolution observations than those 
presented here are needed to fully isolate disks in most embedded systems 
and test detailed theoretical predictions.

\section{Simple Mass Comparison}

How does the circumstellar mass in the optical systems and embedded systems
compare? 
The $\lambda$ = 2.7 mm emission provides a measure of
circumstellar mass in these systems. 
Detailed modeling, including data from a variety of wavelengths, 
is needed to provide the best mass estimates, but the accuracy of any 
mass estimate is limited by the uncertainty in the dust emissivity and 
the true structure of the source.
Since the purpose of this paper is to present a survey, we will apply
a very simple emissivity model, and some assumptions, to 
make a rough mass estimate for the different systems. We adopt a
single-temperature, optically thin source model  so that
F$_{\nu} = B_{\nu}(T_{dust})\kappa_{\nu}M/D^{2}$, where
$B_{\nu}(T)$ is the Planck function, $T_{dust}$ is the temperature of the dust,
$\kappa_{\nu}$ is the dust mass opacity, M is the mass of gas and dust, and 
D is the distance to the source.
For the dust temperature, we will assume characteristic
temperatures of 35~K for the deeply embedded objects
and 60~K for the disk dominated systems.
For the dust mass opacity
(\markcite{b91}Beckwith \& Sargent 1991; \markcite{dra}Draine 1990;
\markcite{poll}Pollack et al. 1994; 
\markcite{stog} Stognienko, Henning, \& Ossenkopf 1995),
we adopt a $\kappa_{\nu}$ which is consistent with other works 
(e.g. \markcite{b91}Beckwith \& Sargent 1991; 
\markcite{ohas}Ohashi et al. 1991; \markcite{ost}Osterloh \& Beckwith 1995):
$\kappa_{\nu}$ = 0.1($\nu$/1200 GHz) cm$^{2}$ g$^{-1}$,
corresponding to $\kappa_{\nu}$ = 0.009 cm$^{2}$ g$^{-1}$ at $\lambda$~=~2.7~mm.
Although we do not expect this simple model to give accurate masses,
it provides rough estimates that are adequate for qualitative
comparisons and are arguably within a factor of 2 of the likely mass.
More detailed modeling of the individual sources will be done in 
subsequent papers.

Table 3 lists the estimated mass for each source, as well as the best
fitted positions from the highest resolution image 
(typical positional uncertainties of 0$\farcs$15).
The masses show more than a factor of a hundred range from
the least to most massive system in our sample.
Where there is overlap, there is reasonable agreement between the simple 
model and published mass estimates.
For example, the mass for HL Tauri in Table 3, 0.05 \msun, is within the
range of masses previously found, 0.05 to 0.1 \msun\ 
(\markcite{b90}Beckwith et al. 1990; \markcite{mund96}Mundy et al. 1996, 
\markcite{wil96}Wilner, Ho, \& Rodr\'{i}guez 1996; \markcite{close}Close et al. 1997), 
and our mass for DG Tauri, 0.03~\msun, is consistent with previous estimates
of 0.02 to 0.04~\msun\ (\markcite{b90}Beckwith et al. 1990; 
\markcite{d96}Dutrey et al. 1996).
For the embedded systems IRAS 16293-2422 and L1448 IRS3, our estimated
masses of 1.1~\msun\ and 0.74~\msun\, respectively, are similar to
other interferometric estimates, 1~\msun\ for IRAS 16293-2422 
(e.g. \markcite{mund92}Mundy et al. 1992)
and 0.5~\msun\ for L1448 IRS3 B 
(e.g. \markcite{ter93}Terebey, Chandler, \& Andr\'{e} 1993). 

The circumstellar masses for the two categories, optical/infrared and
deeply embedded sources, follow the expected trend: the embedded 
objects typically have a factor of 5 or so larger masses.
Comparing circumstellar masses and stellar masses 
for the optical sources, as expected,
these stars have already acquired most of their final mass.
The luminosities of the optical/infrared systems range from 
$\sim$1 to 30 \lsun, suggesting 
central masses of 0.5 to 1.5 \msun, whereas their circumstellar masses in
Table 3 range from 0.01 to 0.08 \msun.

The embedded systems have typical circumstellar masses of $\sim$0.5 \msun,
with the largest one, NGC 1333 IRAS~4~A, near 3 \msun.
The luminosities of the embedded systems range from $\sim$ 1 \lsun\ for 
VLA 1623 to $\sim$ 50 \lsun\ for the NGC 1333 SVS 13 system.
Given the star formation regions in which they are found
(NGC 1333 and Ophiuchi) and their luminosities, 
it is likely that the embedded sources are forming a similar range of 
stellar masses to that of the optical/infrared sources.
Thus, the current circumstellar masses are generally comparable
to the expected stellar masses. Unfortunately, the current masses of the
forming stellar bodies is unknown. Nearly all of the
observed systems have outflows, centralized dust heating, and
significant luminosities, indicating that a stellar object is
present. The current stellar mass required
to generate the observed luminosity can be roughly estimated, 
if it is assumed that
the observed circumstellar mass is accreting on a central
object at a constant rate over the typical duration of this
phase of 10$^{5}$ years.
For a stellar radius of 3~R$_{\sun}$, the derived stellar masses
are in the range of 0.05 \msun\ to 0.4 \msun.
This simple approach suggests that the stellar cores in
the embedded systems are likely to have already
attained a significant fraction (10 to 40\%) 
of their final stellar masses.

\section{Young Multiple Systems}

All of the embedded sources in our survey are either part of small
groupings or are in close binary systems.
Even though our sample may be biased due to our selection criteria,
binary systems appear to be common in the
earliest stages of star formation.
Three possible explanations for the large number of embedded binary systems
are:
(1) small sample statistics,
(2) selection effects in our sample, such that binary,
embedded sources were preferentially chosen, or
(3) true high multiplicity among younger systems
(\markcite{gh93}Ghez, Neugebauer, \& Matthews 1993).

On the first point, even though we observed a small number of systems,
the probability of choosing six multiple systems from a Taurus-like
binary fraction distribution of $\sim$~60\%
(e.g. \markcite{gh93}Ghez, Neugebauer, \&, Matthews 1993)
is low (only 4.7\%) but it can not be ruled out.
On the second point, one obvious selection bias would be if
binary systems typically have more massive envelopes than
single star systems, making embedded binary systems brighter at millimeter
wavelengths.
This supposition is opposite to the trend seen in older,
optical T Tauri binaries;
studies of these systems (\markcite{b90}Beckwith et al. 1990;
\markcite{j94}Jensen, Mathieu, \& Fuller 1994, \markcite{j96}1996;
\markcite{ost}Osterloh \& Beckwith 1995)
provide statistical evidence that T Tauri binary systems have {\it less}
millimeter emission than single systems.
However, the emission properties of young binary systems could be
time dependent--- young embedded binary systems could be brighter
millimeter sources than coeval single star systems, because
they have more massive envelopes.
As they evolve, their envelopes disappear and the remaining material is
more rapidly processed through their circumstellar disks than
in single star systems, and they
become less bright at millimeter wavelengths than comparable single star
systems.
Data on more embedded systems are needed to test this possibility.

On the third point, the evolution of binarity among stellar systems
is an open question. Our results support the idea that
binaries and multiple systems may be more common in younger systems;
however, our criteria for identifying multiple systems is fairly loose.
Systems with separations of 5000 AU or more are valid pairs
for forming stars because the reservoir of cloud material
that the forming stars draws from is typically several
thousand, or more, AU. But, as such systems evolve, the loss of envelope
mass and interactions with other stars forming in the cloud provide
mechanisms for unbinding loose binary systems.
The evidence for evolutionary trends
in the binarity of optical young stars is contradictory.
Speckle observations of the Hyades cluster, a young main-sequence cluster,
find that the occurrence of binary systems in the cluster is larger
than in the local solar neighborhood but less than the Taurus clouds
(\markcite{pat98}Patience et al. 1998), but
studies of additional optical clusters did not confirm the trend
of decreasing binary occurrence with age
(\markcite{pat99}Patience et al. 1999).
It is also possible that binary formation mechanisms have
a dependence on the initial conditions of the pre-collapse cloud
(\markcite{dur}Durisen \& Sterzik 1994) or the density of the star
forming cluster (\markcite{bov}Bouvier, Rigaut, \& Nadeau 1997),
so simple comparison of different regions may not be valid.
Broader survey work is needed to separate the different possible
factors. The solid result from our work is that multiplicity is common
and established early in the formation process and on a variety of
scales.

The most favored mechanism for the early formation of binary and multiple
stellar systems is fragmentation within either the initial cloud core or the
circumstellar disk.
Fragmentation during the earliest stages of the
isothermal collapse of a cloud core, due to perturbations or non-spherical
cores, can form binary systems with separations ranging from 10 to 10$^{4}$ AU
(\markcite{bo79}Boss \& Bodenheimer 1979;
\markcite{mon}Monaghan \& Lattanzio 1986;
\markcite{bon91}Bonnell et al. 1991;
\markcite{bon92}Bonnell \& Bastien 1992;
\markcite{bo93}Boss 1993; \markcite{bat}Bate, Bonnell, \& Price 1995).
Fragmentation due to gravitational instabilities in the circumstellar 
disk systems
may form binary systems with separations ranging from 10 $R_{\sun}$ to 100 AU
(\markcite{a89}Adams, Ruden, \& Shu 1989;
\markcite{shu90}Shu et al. 1990; \markcite{bon94}Bonnell 1994;
\markcite{bon94b}Bonnell \& Bate 1994).
In our survey, the majority of the circumstellar mass in embedded systems
is in the
large-scale envelope, with very little mass in circumstellar disks.
This would suggest that fragmentation occurs during the early
evolution of the core in most systems.
However, the Perseus clouds are distant enough that
we would not detect close ($<$ 150 AU separation) binary systems,
the primary regime of the disk fragmentation and the dominate population
for optical binary stars.
Of the optical sources in our survey, all of which are located in the 
Taurus cloud,
only two appear to be binary systems--- GG Tauri and L1551 IRS5.
Both of these sources could have been formed by disk fragmentation
since they both have circumbinary structures.
Observations with resolution of 10 AU, or smaller, are needed to probe the
fragmentational history of close binaries.

Morphologically, we can identify three types of multiple systems in our
sample: separate envelope, common envelope, and common disk systems
(Table 4).
The characteristics of the different systems are defined by the
broad distribution of the circumstellar material. Separate envelope
systems exhibit clearly distinct centers of gravitational concentration
with separations of $\ge$6000 AU; the components are within a larger
surrounding core of low density material. Common envelope systems have
one primary core of gravitational concentration which breaks into
multiple objects at separations of 100 - 3000 AU. Common disk systems
have separations of $\le$ 100 AU and typically have circumbinary
disk-like distributions of material.  Table 5 lists the binary systems
with our classification, their association, and the projected separation.
The association number assigned in Table 5 identifies members of 
common envelope or common disk systems.

There are several clear connections between these morphological
distinctions and other works. The study of the separation distribution
of optical binaries by \markcite{lar95}Larson (1995) found a knee in the
distribution at 0.04 pc (8250 AU) which was identified with the Jeans size.
Larson suggested that systems on that scale and larger formed by fragmentation
and separate collapse, exactly the structure found in the separate
envelope systems.
This scenario of prompt initial fragmentation is
not new (e.g. \markcite{lar78}Larson 1978, \markcite{pring}Pringle 1989,
\markcite{bon91}Bonnell et al. 1991); it was discussed recently
by \markcite{bon97}Bonnell et al. (1997) in the context of small
cluster formation.
The critical issue is that the collapse is initiated in a system
which contains multiple Jeans masses in a weakly condensed
configuration; one example of such a system would be a prolate Gaussian
distribution with several Jeans masses along the long axis and one Jeans
mass across the short axes.
In systems with large separations, ambipolar diffusion in
self-gravitating, magnetic clouds may play an important role
(e.g. \markcite{mous}Mouschovias 1991).

The common envelope systems can be linked with models for the fragmentation
of moderately centrally-condensed spherical systems
(\markcite{burk}Burkert \& Bodenheimer 1993; \markcite{bo95}Boss 1995,
\markcite{bo97}Boss 1997).
In these models, fragmentation occurs in
the dense central region within an overall single core. The primary requirement
for fragmentation is that the central region has a fairly flat distribution.
However, evidence of this flat region is erased quickly once the fragmentation 
and collapse occurs, and the forming multiple system is left embedded 
within a single centrally condensed core. 
Finally, the common disk systems are similar to models
of high angular momenta systems (\markcite{art}Artymowiez \& Lubow 1994;
\markcite{bat97}Bate \& Bonnell 1997).
This implies that close stellar systems may be created by fragmentation of
early disks or pseudo-disk structures created early in the collapse 
(e.g. \markcite{Galli}Galli \& Shu 1993).
The distribution of material between circumstellar and circumbinary structures
in such systems, depends sensitively on the angular momentum of the infalling
material.

\section{Conclusions}

We have presented the first sub-arcsecond millimeter wavelength 
survey of the dust continuum emission towards 24 young stellar systems.
Morphologically, the systems can be divided into young, embedded objects 
and older optical/infrared sources.
The optical systems show compact emission from circumstellar
disks that only begins to be resolved at $\sim$1$\arcsec$ resolution.
In two systems, HL Tauri and DG Tauri, we resolve the circumstellar disk.
The embedded systems show thermal dust emission that is dominated by 
their circumstellar envelopes with weak residual
emission at small scales.
The embedded systems have $\ge$ 85\% of their $\lambda$~=~2.7~mm 
continuum emission from large scale structures.
This suggests that young circumstellar disks of the embedded systems
are not prominent mass
reservoirs compared to the inner circumstellar envelope.
In fact, for most of the embedded systems, there is no obvious discontinuity
in flux between 1$\arcsec$ and 3$\arcsec$.
Thus, if there are circumstellar disks in these systems, they are not 
more massive than the expected mass of the power-law envelope extrapolated
to small scales.

Simple estimates of the circumstellar mass 
(not including the star) in our systems show that 
the circumstellar mass of the embedded systems (ranging from 0.04 to 2.9 \msun) 
is, as expected,
larger than the circumstellar mass of the optical/infrared systems
(0.01 to 0.08 \msun).
Typically, the embedded systems have a factor of 5 or so larger 
circumstellar masses.
This follows the broad trend that the 
older optical/infrared sources have nearly reached their final stellar
mass, since there is little mass left to accrete, and that the younger,
embedded sources have not reached their final stellar mass, 
since it is expected that a fraction of the circumstellar mass will 
be accreted.
However, simple calculations for the embedded systems
based on luminosity and circumstellar mass, suggest that the 
embedded systems may have already accreted a significant fraction
(10\% or more) of their probable final stellar mass.

The survey has a large number of multiple systems; all of the embedded
objects are in small groupings or binary systems.
Morphologically, based on the system separation, 
we categorize our sample into three types of multiple
systems:  separate envelope (separation $\ge$~6500~AU),
common envelope (separation 150-3000 AU), 
and common disk (separation $\le$ 100 AU).
The separate envelope group links with the idea of 
initial fragmentation and collapse 
of multiple Jeans masses within a weakly condensed configuration 
(e.g. \markcite{lar78}Larson 1978; \markcite{pring}Pringle 1989; 
\markcite{bon91}Bonnell et al. 1991).
The common envelope group is similar to fragmentation of a 
moderately centrally condensed spherical system 
(e.g. \markcite{burk}Burkert \& Bodenheimer 1993; \markcite{bo95}Boss 1995, 
\markcite{bo97}1997).
Finally, the common disk group connects with fragmentation of 
the early disk 
due to high angular momentum (\markcite{art}Artymowiez \& Lubow 1994; 
\markcite{bat97}Bate \& Bonnell 1997).
These three categories suggest three distinct times during the
star formation process where specific fragmentation mechanisms
dominate the binary formation process. 
Thus, the separation morphology may also indicate a time sequence
at which binaries form in stellar evolution.

\acknowledgments
We thank the Hat Creek staff for their efforts in the construction
and operation of the long baselines array.
We especially thank Pedro Safier for discussions on cloud collapse.
We also thank Eve Ostriker and Steve Lubow for useful discussions.
This work was supported by NSF Grants NSF-FD93-20238, NSF-FD96-13716, 
and  AST-9314847.
LWL and LGM acknowledge support from NASA grant NAGW-3066.

\newpage

\begin{deluxetable}{lcccccc}
\tablewidth{0pt}
\tablecaption{Source List}
\tablehead{\colhead{Source}&\colhead{Distance}&\colhead{Optical/IR}&
\colhead{Class}&\colhead{Main}&\colhead{Secondary}&\colhead{Dist.}\\
\colhead{}&\colhead{(pc)}&\colhead{or Embedded}&\colhead{}&
\colhead{Calibrator}&\colhead{Calibrator}&\colhead{Ref.}}
\startdata
L1448 IRS3 A      & 300 & Embedded   & 0  & 3C111    & 0336+323 & 2 \nl
L1448 IRS3 B      & 300 & Embedded   & 0  & 3C111    & 0336+323 & 2 \nl
L1448 IRS3 C      & 300 & Embedded   & 0  & 3C111    & 0336+323 & 2 \nl
NGC1333 IRAS2 A   & 350 & Embedded   & 0  & 3C111    & 0336+323 & 2 \nl
NGC1333 IRAS2 B   & 350 & Embedded   & 0  & 3C111    & 0336+323 & 2 \nl
SVS 13 A1         & 350 & Optical/IR & \nodata  & 3C111    & 0336+323 & 2 \nl
SVS 13 A2         & 350 & Embedded   & \nodata  & 3C111    & 0336+323 & 2 \nl
SVS 13 B          & 350 & Embedded   & \nodata  & 3C111    & 0336+323 & 2 \nl
SVS 13 C          & 350 & Embedded   & \nodata  & 3C111    & 0336+323 & 2 \nl
NGC1333 IRAS4 A1  & 350 & Embedded   & 0  & 3C111    & 0336+323 & 2 \nl
NGC1333 IRAS4 A2  & 350 & Embedded   & 0  & 3C111    & 0336+323 & 2 \nl
NGC1333 IRAS4 B   & 350 & Embedded   & 0  & 3C111    & 0336+323 & 2 \nl
NGC1333 IRAS4 C   & 350 & Embedded   & \nodata  & 3C111    & 0336+323 & 2 \nl
DG Tauri          & 140 & Optical/IR & II & 3C111    & 0431+206 & 1 \nl
DG Tauri B \tablenotemark{a}   & 140 & Optical/IR & I  & 3C111    & 0431+206 & 1 \nl
L1551 IRS5        & 140 & Optical/IR & I  & 3C111    & 0336+323 & 1 \nl
HL Tauri          & 140 & Optical/IR & II & 0530+135 & 0431+206 & 1 \nl
GG Tauri          & 140 & Optical/IR & II & 0530+135 & 0431+206 & 1 \nl
GM Aurigae        & 140 & Optical/IR & II & 3C111    &  3C123   & 1 \nl
VLA 1623          & 160 & Embedded   & 0  & 1733-130 & 1625-254 & 3 \nl
IRAS 16293-2422 A & 160 & Embedded   & 0  & 1733-130 & 1625-254 & 3 \nl
IRAS 16293-2422 B & 160 & Embedded   & 0  & 1733-130 & 1625-254 & 3 \nl
\tablenotetext{a}{DG Tauri B observed near the FWHM of primary beam.}
\tablerefs{(1) \markcite{elias}Elias 1978; (2) \markcite{herb}Herbig \& Jones 1983; (3) Whittet 1974}
\enddata
\end{deluxetable}

\newpage

\begin{deluxetable}{llccc}
\tablewidth{0pt}
\tablecaption{Source Flux}
\tablehead{\colhead{Source}&\colhead{Panel}&\colhead{Peak Flux}&\colhead{Integrated Flux}&
\colhead{Box Size}\\
\colhead{}&\colhead{}&\colhead{mJy/beam}&\colhead{mJy}}
\startdata
L1448~IRS3~A    &(a)&  26.5$\pm$1.6&  23.1$\pm$ 2.6&11$\farcs2$ $\times$  6$\farcs3$      \nl  
		&(b)&  14.5$\pm$1.5&  26.7$\pm$ 3.3& 8$\farcs3$ $\times$  5$\farcs8$      \nl
		&(c)&   6.8$\pm$1.6&  19.3$\pm$ 3.7& 2$\farcs3$ $\times$  2$\farcs9$      \nl
		&(d)&  $<$6.7      &  \nodata      &  \nodata                              \nl
L1448~IRS3~B    &(a)& 101.5$\pm$1.6& 134.6$\pm$ 3.9&17$\farcs0$ $\times$  9$\farcs8$      \nl  
		&(b)&  84.9$\pm$1.5& 135.6$\pm$ 4.8&11$\farcs0$ $\times$  9$\farcs0$      \nl  
		&(c)&  41.1$\pm$1.6& 135.2$\pm$ 6.7& 4$\farcs5$ $\times$  4$\farcs9$      \nl
		&(d)&  22.5$\pm$2.3& 115.7$\pm$ 9.5& 2$\farcs8$ $\times$  2$\farcs4$      \nl
L1448~IRS3~C    &(a)&  14.9$\pm$1.6&  31.7$\pm$ 4.1&10$\farcs7$ $\times$ 17$\farcs0$      \nl
		&(b)&  11.7$\pm$1.5&  31.9$\pm$ 3.7& 5$\farcs0$ $\times$ 12$\farcs0$      \nl
		&(c)&   9.8$\pm$1.6&  14.3$\pm$ 3.2& 2$\farcs5$ $\times$  2$\farcs0$      \nl
		&(d)&   8.7$\pm$2.3&   8.7$\pm$ 2.3& 0$\farcs7$ $\times$  0$\farcs5$      \nl
N1333 IRAS2A    &(a)&  46.5$\pm$1.3&  82.8$\pm$ 4.0&16$\farcs4$ $\times$ 16$\farcs3$      \nl
		&(b)&  36.2$\pm$1.2&  74.4$\pm$ 4.0&11$\farcs0$ $\times$ 12$\farcs0$      \nl
		&(c)&  22.3$\pm$1.7&  36.1$\pm$ 4.4& 2$\farcs9$ $\times$  2$\farcs2$      \nl
		&(d)&  18.4$\pm$2.7&  22.4$\pm$ 4.8& 1$\farcs4$ $\times$  0$\farcs9$      \nl
N1333 IRAS2B    &(a)&  21.3$\pm$1.3&  27.7$\pm$ 3.2&12$\farcs8$ $\times$ 13$\farcs0$      \nl
                &(b)&  19.6$\pm$1.2&  24.4$\pm$ 2.7& 6$\farcs6$ $\times$  9$\farcs2$      \nl
		&(c)&  18.9$\pm$1.7&  24.7$\pm$ 3.5& 1$\farcs9$ $\times$  2$\farcs2$      \nl
		&(d)&  16.9$\pm$2.7&  24.3$\pm$ 5.1& 1$\farcs2$ $\times$  1$\farcs2$      \nl
SVS 13 A        &(a)&  47.3$\pm$1.1& 101.3$\pm$ 4.2&26$\farcs0$ $\times$ 15$\farcs0$      \nl
		&(b)&  37.4$\pm$1.1& 100.3$\pm$ 4.7&20$\farcs7$ $\times$ 10$\farcs2$      \nl
		&(c)&  19.2$\pm$1.5&  38.7$\pm$ 4.1& 2$\farcs9$ $\times$  2$\farcs9$      \nl
		&(d)&  11.0$\pm$2.2&  38.0$\pm$ 6.6& 1$\farcs7$ $\times$  2$\farcs2$      \nl
SVS 13 B        &(a)&  52.0$\pm$1.1& 123.0$\pm$ 4.5&31$\farcs5$ $\times$ 14$\farcs0$      \nl
		&(b)&  41.7$\pm$1.1& 110.4$\pm$ 3.7&10$\farcs1$ $\times$ 13$\farcs0$      \nl
		&(c)&  25.3$\pm$1.5&  41.4$\pm$ 3.6& 2$\farcs2$ $\times$  3$\farcs0$      \nl
		&(d)&  19.4$\pm$2.2&  48.2$\pm$ 6.6& 1$\farcs5$ $\times$  2$\farcs5$      \nl
SVS 13 C        &(a)&  11.7$\pm$1.1&  21.0$\pm$ 2.5&14$\farcs0$ $\times$ 10$\farcs0$      \nl
		&(b)&   9.6$\pm$1.1&  19.8$\pm$ 2.7& 9$\farcs1$ $\times$  7$\farcs6$      \nl
		&(c)&   8.5$\pm$1.5&   8.5$\pm$ 1.5& 1$\farcs1$ $\times$  1$\farcs0$      \nl
		&(d)&  11.1$\pm$2.2&  11.1$\pm$ 2.2& 0$\farcs7$ $\times$  0$\farcs5$      \nl
N1333 IRAS4A    &(a)& 351.2$\pm$3.1& 544.2$\pm$13.6&25$\farcs0$ $\times$ 24$\farcs0$      \nl
		&(b)& 280.4$\pm$1.9& 525.6$\pm$ 9.2&12$\farcs0$ $\times$ 18$\farcs5$      \nl
		&(c)& 172.2$\pm$2.1& 449.7$\pm$ 9.8& 5$\farcs4$ $\times$  6$\farcs2$      \nl
~~A1 Only 	&(d)& 107.0$\pm$2.9& 324.1$\pm$12.0& 2$\farcs9$ $\times$  2$\farcs2$      \nl
~~A2 Only      &(d)&  23.0$\pm$2.9&  81.2$\pm$ 8.1& 1$\farcs8$ $\times$  1$\farcs6$      \nl
\tablebreak
N1333 IRAS4B    &(a)& 143.3$\pm$3.1& 180.3$\pm$ 7.9&12$\farcs0$ $\times$ 17$\farcs0$      \nl
		&(b)& 129.1$\pm$1.9& 172.1$\pm$ 6.0& 8$\farcs5$ $\times$ 11$\farcs0$      \nl
		&(c)&  94.0$\pm$2.1& 148.9$\pm$ 5.9& 3$\farcs4$ $\times$  3$\farcs6$      \nl
		&(d)&  57.6$\pm$2.9& 128.8$\pm$ 7.9& 1$\farcs7$ $\times$  1$\farcs6$      \nl
N1333 IRAS4C    &(a)&  47.8$\pm$3.1&  49.8$\pm$ 5.5& 9$\farcs0$ $\times$ 11$\farcs0$      \nl
                &(b)&  48.5$\pm$1.9&  50.7$\pm$ 3.8& 5$\farcs5$ $\times$  7$\farcs0$      \nl
		&(c)&  39.9$\pm$2.1&  57.0$\pm$ 4.9& 3$\farcs0$ $\times$  2$\farcs8$      \nl
		&(d)&  26.9$\pm$2.9&  51.6$\pm$ 6.9& 1$\farcs4$ $\times$  1$\farcs5$      \nl
DG Tauri        &(a)&  57.7$\pm$2.7&  66.0$\pm$ 5.8&11$\farcs4$ $\times$ 11$\farcs2$      \nl
                &(b)&  53.6$\pm$2.0&  73.8$\pm$ 6.2& 9$\farcs0$ $\times$ 10$\farcs4$      \nl
                &(c)&  46.0$\pm$1.9&  71.3$\pm$ 4.8& 2$\farcs9$ $\times$  2$\farcs9$      \nl
		&(d)&  34.6$\pm$1.6&  68.9$\pm$ 5.1& 2$\farcs1$ $\times$  2$\farcs5$      \nl
DG Tauri B \tablenotemark{a}     &(a)&  45.0$\pm$4.8&  78.4$\pm$11.3&13$\farcs0$ $\times$ 12$\farcs0$      \nl
		&(b)&  38.8$\pm$3.5&  72.7$\pm$10.9& 7$\farcs8$ $\times$ 11$\farcs8$      \nl
		&(c)&  30.6$\pm$3.4&  47.8$\pm$ 7.2& 1$\farcs9$ $\times$  3$\farcs1$      \nl
		&(d)&  22.7$\pm$2.8&  49.4$\pm$ 8.8& 1$\farcs4$ $\times$  1$\farcs8$      \nl
L1551~IRS5      &(a)& 133.9$\pm$2.6& 173.3$\pm$ 7.5&17$\farcs0$ $\times$ 14$\farcs0$      \nl
		&(b)& 120.7$\pm$2.5& 177.2$\pm$ 7.9&12$\farcs0$ $\times$  8$\farcs8$      \nl
		&(c)&  77.9$\pm$3.3& 145.2$\pm$ 9.1& 2$\farcs7$ $\times$  3$\farcs5$      \nl 
		&(d)&  56.0$\pm$3.9& 107.0$\pm$11.1& 1$\farcs3$ $\times$  1$\farcs9$      \nl
HL Tauri        &(a)& 102.7$\pm$1.7& 108.6$\pm$ 4.6&19$\farcs0$ $\times$ 17$\farcs8$      \nl
                &(b)&  90.9$\pm$1.7& 113.6$\pm$ 4.8& 9$\farcs0$ $\times$  9$\farcs6$      \nl
                &(c)&  70.3$\pm$2.4& 106.2$\pm$ 6.0& 2$\farcs6$ $\times$  2$\farcs9$      \nl
		&(d)&  48.8$\pm$2.9& 106.9$\pm$ 7.8& 1$\farcs8$ $\times$  1$\farcs5$      \nl
GG Tauri        &(a)&  56.7$\pm$1.8&  73.5$\pm$ 4.4&12$\farcs6$ $\times$ 12$\farcs4$      \nl
		&(b)&  27.2$\pm$1.2&  72.5$\pm$ 3.6& 7$\farcs5$ $\times$  6$\farcs7$      \nl
                &(c)&  10.3$\pm$1.2&  78.0$\pm$ 5.0& 4$\farcs8$ $\times$  4$\farcs9$      \nl 
		&(d)&  10.8$\pm$1.5&  95.2$\pm$ 6.8& 4$\farcs4$ $\times$  4$\farcs5$      \nl
GM Aurigae      &(a)&  20.3$\pm$1.1&  22.0$\pm$ 2.6&12$\farcs0$ $\times$ 14$\farcs0$      \nl
		&(b)&  19.2$\pm$0.9&  22.3$\pm$ 2.0& 7$\farcs8$ $\times$  6$\farcs2$      \nl
		&(c)&  13.6$\pm$1.6&  19.6$\pm$ 3.0& 2$\farcs0$ $\times$  2$\farcs2$      \nl
		&(d)&  13.4$\pm$2.5&  13.4$\pm$ 2.5& 0$\farcs6$ $\times$  0$\farcs5$      \nl 
VLA 1623 A\&B   &(a)&  54.2$\pm$3.0&  72.1$\pm$ 6.8& 9$\farcs4$ $\times$ 18$\farcs0$      \nl
		&(b)&  44.2$\pm$2.2&  53.5$\pm$ 4.3& 5$\farcs4$ $\times$  8$\farcs0$      \nl 
~~A Only       &(c)&  22.8$\pm$2.0&  34.4$\pm$ 4.3& 1$\farcs6$ $\times$  3$\farcs4$      \nl
~~B Only       &(c)&  25.0$\pm$2.0&  32.5$\pm$ 4.3& 1$\farcs6$ $\times$  3$\farcs4$      \nl
~~A Only       &(d)&  22.4$\pm$3.5&  25.5$\pm$ 6.3& 0$\farcs9$ $\times$  1$\farcs7$      \nl
~~B Only       &(d)&  25.8$\pm$3.5&  25.8$\pm$ 3.5& 0$\farcs4$ $\times$  0$\farcs9$      \nl
\tablebreak
IRAS 16293-2422 &(a)& 412.6$\pm$5.8&1017.9$\pm$26.5&22$\farcs2$ $\times$ 27$\farcs2$      \nl
~~A Only       &(b)& 176.4$\pm$4.2& 441.2$\pm$14.1&12$\farcs1$ $\times$ 10$\farcs0$      \nl
~~B Only       &(b)& 305.3$\pm$4.2& 551.4$\pm$14.1&12$\farcs5$ $\times$  9$\farcs7$      \nl
~~A Only       &(c)&  60.1$\pm$4.1& 358.3$\pm$18.5& 5$\farcs2$ $\times$  5$\farcs1$      \nl 
~~B Only       &(c)& 154.1$\pm$4.1& 498.4$\pm$17.2& 6$\farcs0$ $\times$  3$\farcs8$      \nl
~~A Only       &(d)&  43.6$\pm$4.8& 276.2$\pm$22.9& 3$\farcs1$ $\times$  4$\farcs8$      \nl
~~B Only       &(d)& 112.7$\pm$4.8& 424.2$\pm$24.2& 3$\farcs4$ $\times$  4$\farcs9$      \nl
\tablenotetext{a}{DG Tauri B fluxes were corrected for primary beam attenuation; thus fluxes 
given have a larger overall uncertainty than the rest of the survey.}
\enddata
\end{deluxetable}
\newpage

\begin{deluxetable}{lcccc}
\tablewidth{0pt}
\tablecaption{Positions, Simple Estimates of Mass, 
and 5k$\lambda$/50k$\lambda$ Ratio}
\tablehead{\colhead{Source}&\colhead{$\alpha$ (J2000)}&
\colhead{$\delta$ (J2000)}&\colhead{Mass}&
\colhead{5k$\lambda$/50k$\lambda$}\\
\colhead{}&\colhead{}&\colhead{}&\colhead{(\msun)}&\colhead{Ratio}}
\startdata
L1448 IRS3 A       &$ 03^{h}25^{m}36\fs532$ &$ +30^{\arcdeg}45^{\arcmin}21\farcs35$ & 0.09                   & 3.7$\pm$0.4 \nl
L1448 IRS3 B       &$ 03^{h}25^{m}36\fs339$ &$ +30^{\arcdeg}45^{\arcmin}14\farcs94$ & 0.52                   & \nodata     \nl
L1448 IRS3 C       &$ 03^{h}25^{m}35\fs653$ &$ +30^{\arcdeg}45^{\arcmin}34\farcs20$ & 0.12                   & \nodata     \nl
NGC1333 IRAS2 A    &$ 03^{h}28^{m}55\fs571$ &$ +31^{\arcdeg}14^{\arcmin}37\farcs22$ & 0.44                   & 2.5$\pm$0.4 \nl
NGC1333 IRAS2 B    &$ 03^{h}28^{m}57\fs349$ &$ +31^{\arcdeg}14^{\arcmin}15\farcs93$ & 0.14                   & 1.1$\pm$0.2 \nl
SVS 13 A1          &$ 03^{h}29^{m}03\fs750$ &$ +31^{\arcdeg}16^{\arcmin}03\farcs95$ & 0.54 \tablenotemark{a} & 3.3$\pm$0.6 \tablenotemark{a} \nl
SVS 13 A2          &$ 03^{h}29^{m}03\fs374$ &$ +31^{\arcdeg}16^{\arcmin}01\farcs87$ & 0.54 \tablenotemark{a} & 3.3$\pm$0.6 \tablenotemark{a} \nl
SVS 13 B           &$ 03^{h}29^{m}03\fs056$ &$ +31^{\arcdeg}15^{\arcmin}51\farcs67$ & 0.65                   & 2.6$\pm$0.4 \nl
SVS 13 C           &$ 03^{h}29^{m}01\fs951$ &$ +31^{\arcdeg}15^{\arcmin}38\farcs27$ & 0.11                   & \nodata     \nl
NGC1333 IRAS4 A1   &$ 03^{h}29^{m}10\fs510$ &$ +31^{\arcdeg}13^{\arcmin}31\farcs01$ & 2.88 \tablenotemark{a} & 2.8$\pm$0.1 \tablenotemark{a} \nl
NGC1333 IRAS4 A2   &$ 03^{h}29^{m}10\fs413$ &$ +31^{\arcdeg}13^{\arcmin}32\farcs20$ & 2.88 \tablenotemark{a} & 2.8$\pm$0.1 \tablenotemark{a} \nl
NGC1333 IRAS4 B    &$ 03^{h}29^{m}11\fs988$ &$ +31^{\arcdeg}13^{\arcmin}08\farcs10$ & 0.96                   & 1.8$\pm$0.1 \nl
NGC1333 IRAS4 C    &$ 03^{h}29^{m}12\fs813$ &$ +31^{\arcdeg}13^{\arcmin}06\farcs97$ & 0.26                   & 1.0$\pm$0.1 \nl
DG Tauri           &$ 04^{h}27^{m}04\fs686$ &$ +26^{\arcdeg}06^{\arcmin}16\farcs14$ & 0.03                   & 0.9$\pm$0.1 \nl
DG Tauri B         &$ 04^{h}27^{m}02\fs562$ &$ +26^{\arcdeg}05^{\arcmin}30\farcs53$ & 0.04 \tablenotemark{b} & 2.2$\pm$0.4 \nl 
L1551 IRS5 A       &$ 04^{h}31^{m}34\fs143$ &$ +18^{\arcdeg}08^{\arcmin}05\farcs09$ & 0.08 \tablenotemark{a} & 1.8$\pm$0.1 \tablenotemark{a} \nl
L1551 IRS5 B       &$ 04^{h}31^{m}34\fs141$ &$ +18^{\arcdeg}08^{\arcmin}04\farcs74$ & 0.08 \tablenotemark{a} & 1.8$\pm$0.1 \tablenotemark{a} \nl
HL Tauri           &$ 04^{h}31^{m}38\fs413$ &$ +18^{\arcdeg}13^{\arcmin}57\farcs61$ & 0.05                   & 1.3$\pm$0.1 \nl 
GG Tauri           &$ 04^{h}32^{m}30\fs322$ &$ +17^{\arcdeg}31^{\arcmin}40\farcs65$ & 0.03                   & 4.0$\pm$0.6 \nl 
GM Aurigae         &$ 04^{h}55^{m}10\fs983$ &$ +30^{\arcdeg}21^{\arcmin}59\farcs37$ & 0.01                   & 1.1$\pm$0.2 \nl 
VLA 1623 A         &$ 16^{h}26^{m}26\fs396$ &$ -24^{\arcdeg}24^{\arcmin}30\farcs45$ & 0.04                   & 2.4$\pm$0.3 \tablenotemark{a} \nl
VLA 1623 B         &$ 16^{h}26^{m}26\fs318$ &$ -24^{\arcdeg}24^{\arcmin}30\farcs12$ & 0.04                   & 2.4$\pm$0.3 \tablenotemark{a} \nl
IRAS 16293-2422 A  &$ 16^{h}32^{m}22\fs869$ &$ -24^{\arcdeg}28^{\arcmin}36\farcs11$ & 0.49                   & 7.4$\pm$0.4 \nl
IRAS 16293-2422 B  &$ 16^{h}32^{m}22\fs624$ &$ -24^{\arcdeg}28^{\arcmin}32\farcs20$ & 0.61                   & 3.5$\pm$0.1 \nl
\tablenotetext{a}{Close binary systems whose values include both systems.}
\tablenotetext{b}{DG Tauri B observed at FWHM point of primary beam; thus masses given have a
larger uncertainty than the rest of the survey.}
\tablenotetext{}{Note: The mass scales roughly linear with the assumed temperature.
For example, in the case of the embedded objects, M(T) = M(35K)$\times$(35K/T).}
\enddata
\end{deluxetable}
\clearpage

\begin{deluxetable}{lr}
\tablewidth{0pt}\tablecaption{Characteristic Scales for Multiplicity}
\tablehead{\colhead{Classification}&\colhead{Scale (AU)}}
\startdata
Separate Envelope     &     $\ge$ 6000   \nl
Common Envelope       &     150 - 3000   \nl
Common Disk           &     $\le$ 100    \nl
\enddata
\end{deluxetable}
\clearpage

\begin{deluxetable}{llcrr}
\tablewidth{0pt}
\tablecaption{Multiple System Morphology}
\tablehead{\colhead{Source}&\colhead{Type}&\colhead{Assoc.}&\multicolumn{2}{c}{Separation}\\
\colhead{}&\colhead{}&\colhead{}&\colhead{Arcsec}&\colhead{AU}}
\startdata
L1448 IRS3 A       & common envelope    & 1 &  6$\farcs$87 & 2404 \nl
L1448 IRS3 B       & common envelope    & 1 &  6$\farcs$87 & 2404 \nl
L1448 IRS3 C       & separate envelope  &   & 17$\farcs$13 & 5995 \nl
NGC1333 IRAS2 A    & separate envelope  &   & 31$\farcs$20 & 10920\nl
NGC1333 IRAS2 B    & separate envelope  &   & 31$\farcs$20 & 10920\nl
SVS 13 A1          & common envelope    & 2,3 &  5$\farcs$25 & 1838 \nl
SVS 13 A2          & common envelope    & 2,3 &  5$\farcs$25 & 1838 \nl
SVS 13 B           & common envelope    & 3 & 10$\farcs$98 & 3843 \nl
SVS 13 C           & separate envelope  &   & 19$\farcs$50 & 6825 \nl
NGC1333 IRAS4 A1   & common envelope    & 4 &  1$\farcs$72 & 602  \nl
NGC1333 IRAS4 A2   & common envelope    & 4 &  1$\farcs$72 & 602  \nl
NGC1333 IRAS4 B    & separate envelope  &   & 29$\farcs$74 & 10409\nl
NGC1333 IRAS4 C    & separate envelope  &   & 10$\farcs$64 & 3724 \nl
DG Tauri           & separate envelope  &   & 54$\farcs$85 & 7540 \nl
DG Tauri B         & separate envelope  &   & 54$\farcs$85 & 7540 \nl
L1551 IRS5 A       & common disk        & 5 &  0$\farcs$35 & 49   \nl
L1551 IRS5 B       & common disk        & 5 &  0$\farcs$35 & 49   \nl
GG Tauri           & common disk        & 6 &  0$\farcs$25 & 35   \nl
VLA 1623 A         & common envelope    & 7 &  1$\farcs$11 & 178  \nl
VLA 1623 B         & common envelope    & 7 &  1$\farcs$11 & 178  \nl
IRAS 16293-2422 A  & common envelope    & 8 &  5$\farcs$14 & 822  \nl
IRAS 16293-2422 B  & common envelope    & 8 &  5$\farcs$14 & 822  \nl
\enddata
\tablenotetext{a}{Column 3 indicates members of a multiple system.}
\end{deluxetable}
\clearpage

\begin{figure}
\includegraphics{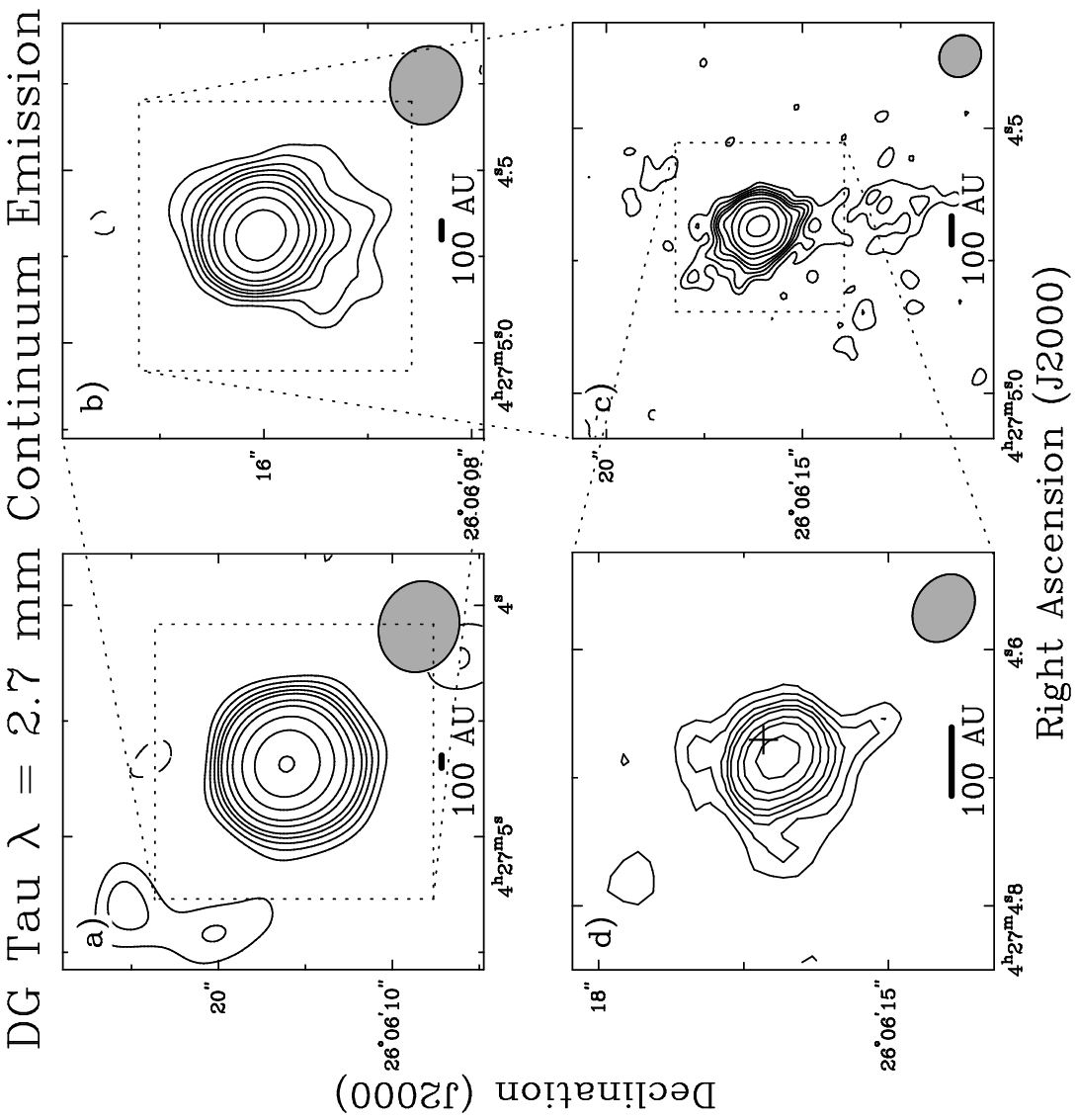}
\vspace{4.5in}
\caption{
DG Tauri maps of the $\lambda$ = 2.7 mm continuum emission.
All panels are contoured in steps of (-4 -3 -2 2 3 4 5 6 8 10 14.14 20 28.28)
$\times$ a rms noise of 2.0 mJy/beam.
(a) $\sigma$ = 2.7 mJy/beam; beam is 5$\farcs$37 $\times$ 4$\farcs$57 P.A. =  72$\arcdeg$.
(b) $\sigma$ = 2.0 mJy/beam; beam is 3$\farcs$12 $\times$ 2$\farcs$72 P.A. =  68$\arcdeg$.
(c) $\sigma$ = 1.9 mJy/beam; beam is 1$\farcs$12 $\times$ 1$\farcs$02 P.A. =  45$\arcdeg$.
(d) $\sigma$ = 1.6 mJy/beam; beam is 0$\farcs$76 $\times$ 0$\farcs$58 P.A. =  56$\arcdeg$.
The cross in panel (d) is the $\lambda$ = 6 cm peak from
Bieging, Cohen, \& Schwartz (1984).}
\end{figure}
\clearpage

\begin{figure}
\includegraphics{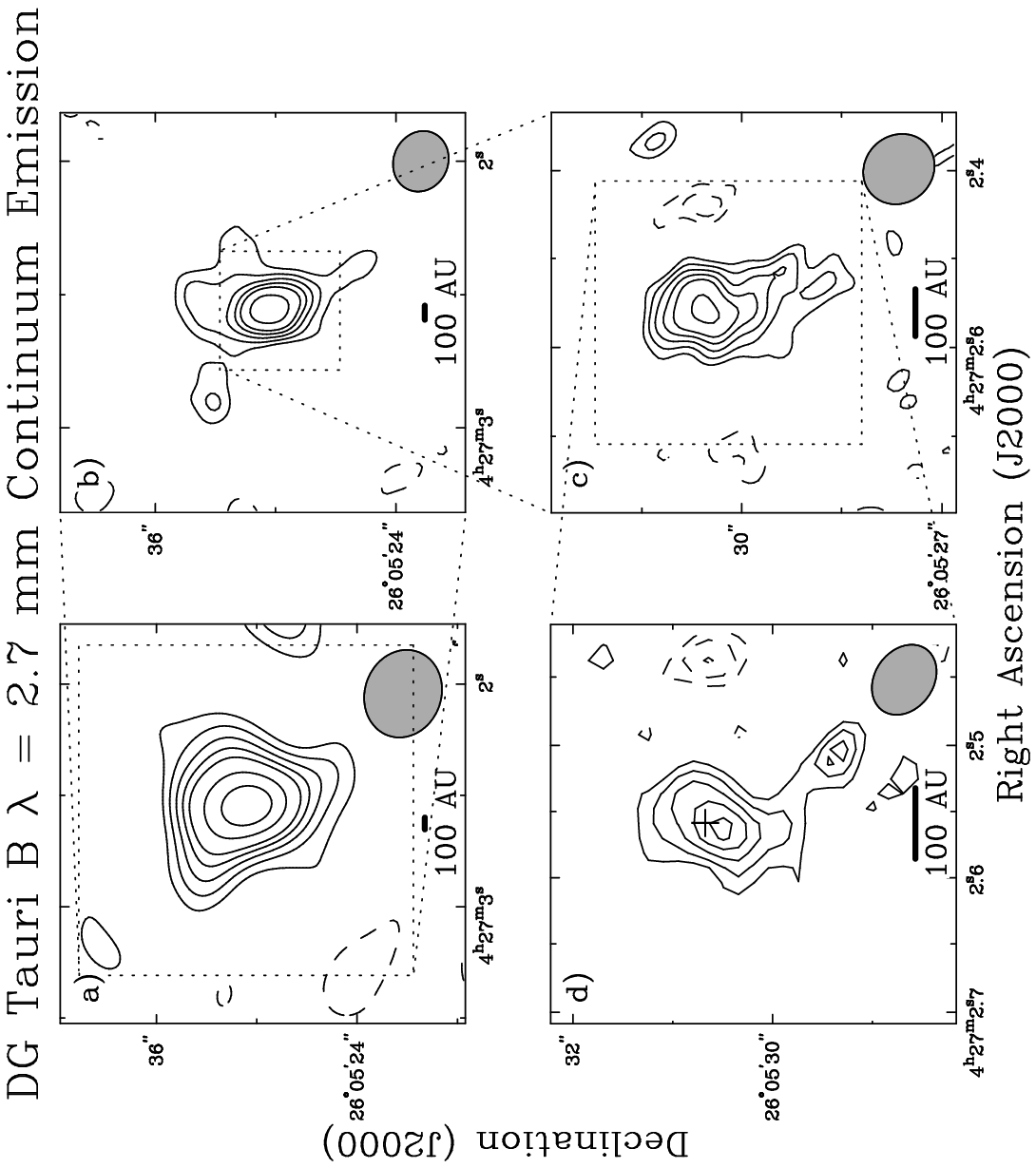}
\vspace{4.5in}
\caption{
DG Tauri B maps of the $\lambda$ = 2.7 mm continuum emission.
All panels are contoured in steps of (-4 -3 -2 2 3 4 5 6 8 10 14.14 20 28.28) 
$\times$ a rms noise of 2.0 mJy/beam.
(a) $\sigma$ = 2.7 mJy/beam; beam is 5$\farcs$37 $\times$ 4$\farcs$57 P.A. =  72$\arcdeg$.
(b) $\sigma$ = 2.0 mJy/beam; beam is 3$\farcs$12 $\times$ 2$\farcs$72 P.A. =  68$\arcdeg$.
(c) $\sigma$ = 1.9 mJy/beam; beam is 1$\farcs$12 $\times$ 1$\farcs$02 P.A. =  45$\arcdeg$.
(d) $\sigma$ = 1.6 mJy/beam; beam is 0$\farcs$76 $\times$ 0$\farcs$58 P.A. =  56$\arcdeg$.
The cross in panel (d) is the $\lambda$ = 3.6 cm peak from Rodr\'{i}guez, Anglada,
\& Raga (1995).}
\end{figure}
\clearpage

\begin{figure}
\includegraphics{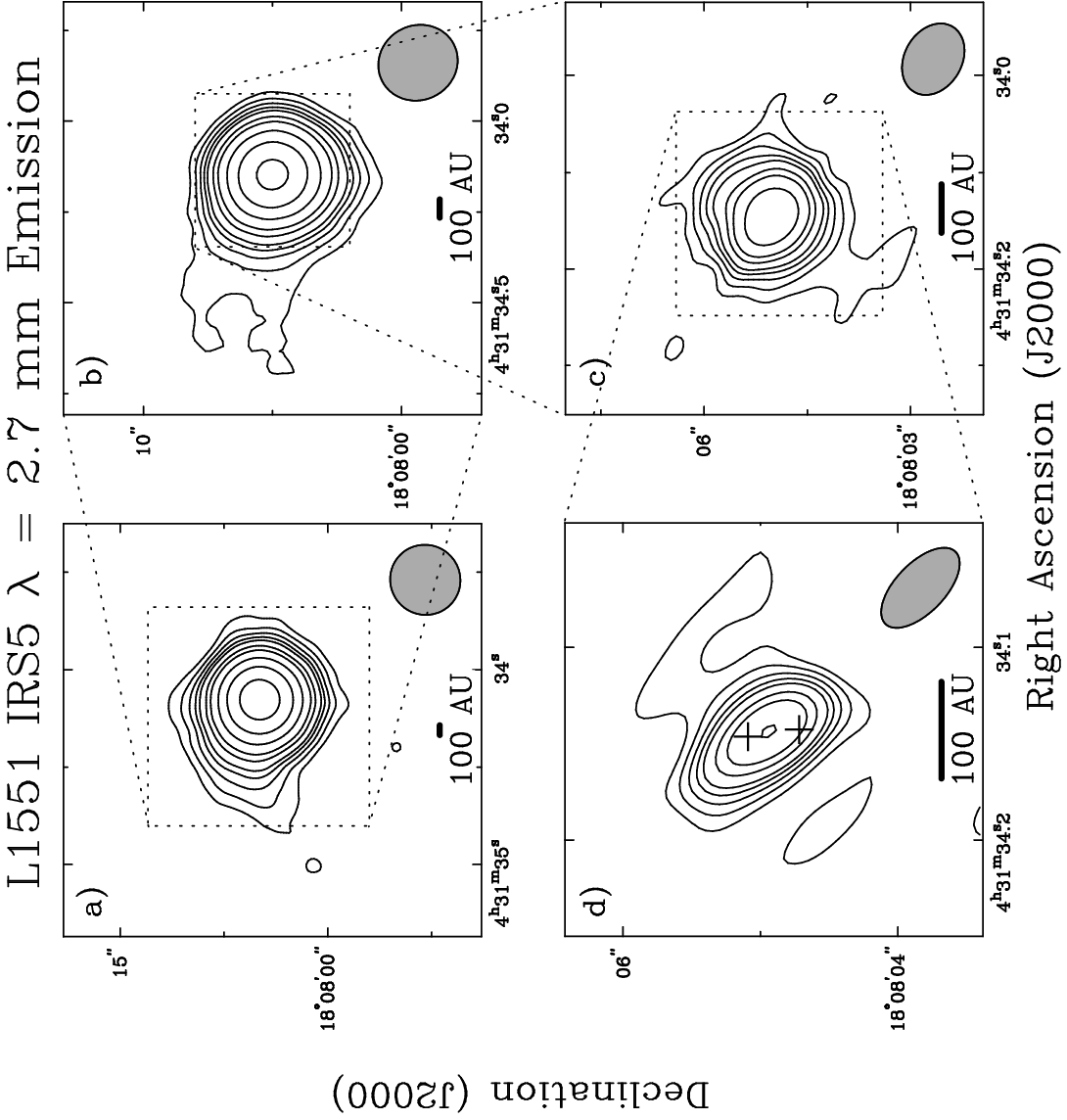}
\vspace{4.5in}
\caption{
L1551 IRS5 maps of the $\lambda$ = 2.7 mm continuum emission.
All panels are contoured in steps of (-4 -3 -2 2 3 4 5 6 8 10 14.14 20 28.28) 
$\times$ a rms noise of 3.9 mJy/beam.
(a) $\sigma$ = 2.6 mJy/beam; beam is 5$\farcs$15 $\times$ 5$\farcs$05 P.A. = -62$\arcdeg$.
(b) $\sigma$ = 2.5 mJy/beam; beam is 3$\farcs$13 $\times$ 2$\farcs$92 P.A. =  31$\arcdeg$.
(c) $\sigma$ = 3.3 mJy/beam; beam is 1$\farcs$11 $\times$ 0$\farcs$85 P.A. =  62$\arcdeg$.
(d) $\sigma$ = 3.9 mJy/beam; beam is 0$\farcs$74 $\times$ 0$\farcs$36 P.A. =  46$\arcdeg$.
The two crosses in panel (d) are the $\lambda$ = 1.3 cm peaks from 
Koerner \& Sargent (1999).}
\end{figure}
\clearpage

\begin{figure}
\includegraphics{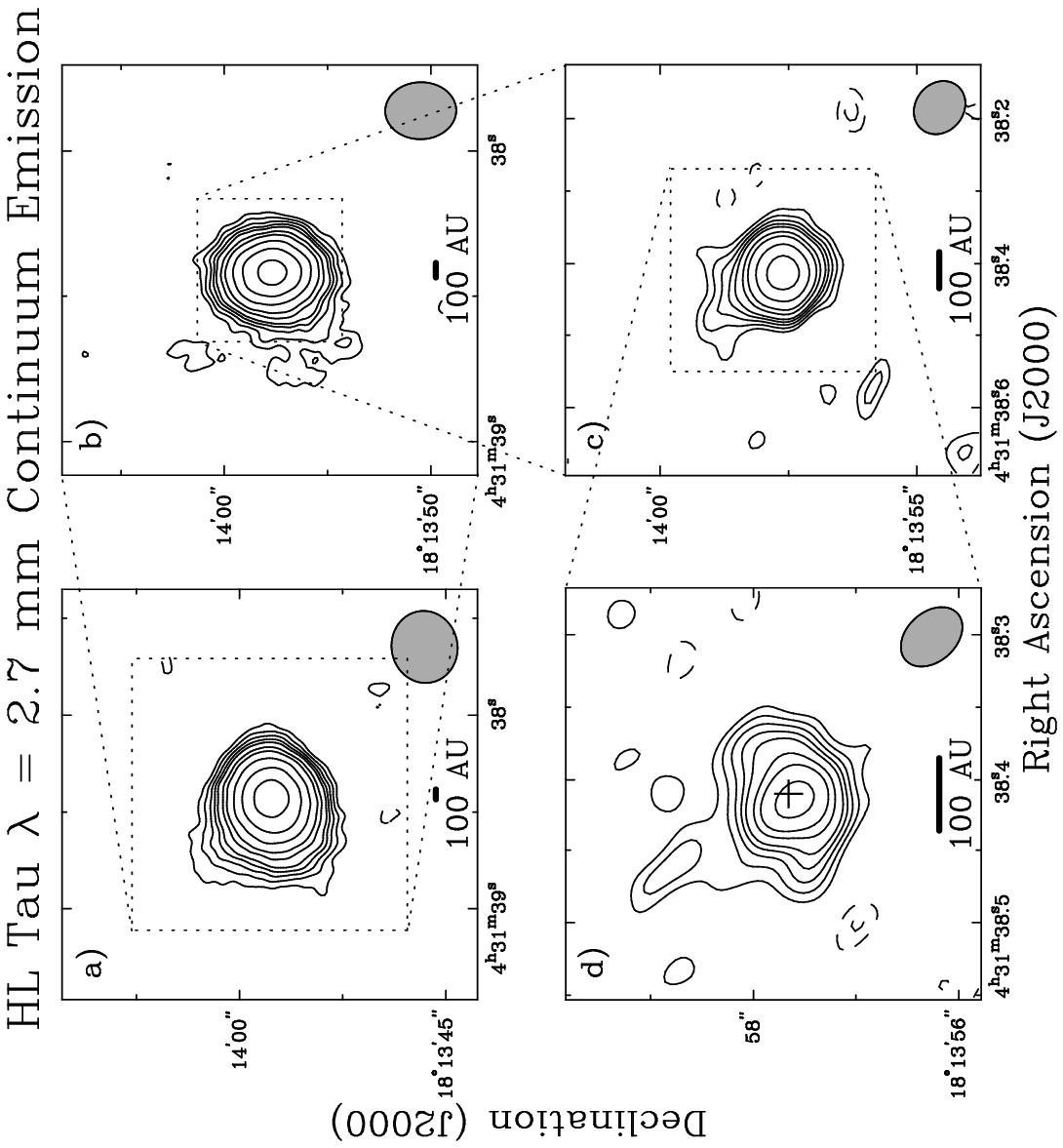}
\vspace{4.5in}
\caption{
HL Tauri maps of the $\lambda$ = 2.7 mm continuum emission.
All panels are contoured in steps of (-4 -3 -2 2 3 4 5 6 8 10 14.14 20 28.28)  
$\times$ a rms noise of 2.9 mJy/beam.
(a) $\sigma$ = 1.7 mJy/beam; beam is 5$\farcs$31 $\times$ 4$\farcs$79 P.A. = -81$\arcdeg$.
(b) $\sigma$ = 1.7 mJy/beam; beam is 3$\farcs$43 $\times$ 2$\farcs$79 P.A. =   1$\arcdeg$.
(c) $\sigma$ = 2.4 mJy/beam; beam is 1$\farcs$11 $\times$ 0$\farcs$94 P.A. =  53$\arcdeg$.
(d) $\sigma$ = 2.9 mJy/beam; beam is 0$\farcs$68 $\times$ 0$\farcs$48 P.A. =  43$\arcdeg$.
The cross in panel (d) is the $\lambda$ = 3.6 cm peak from Rodr\'{i}guez et al. (1994).}
\end{figure}
\clearpage

\begin{figure}
\includegraphics{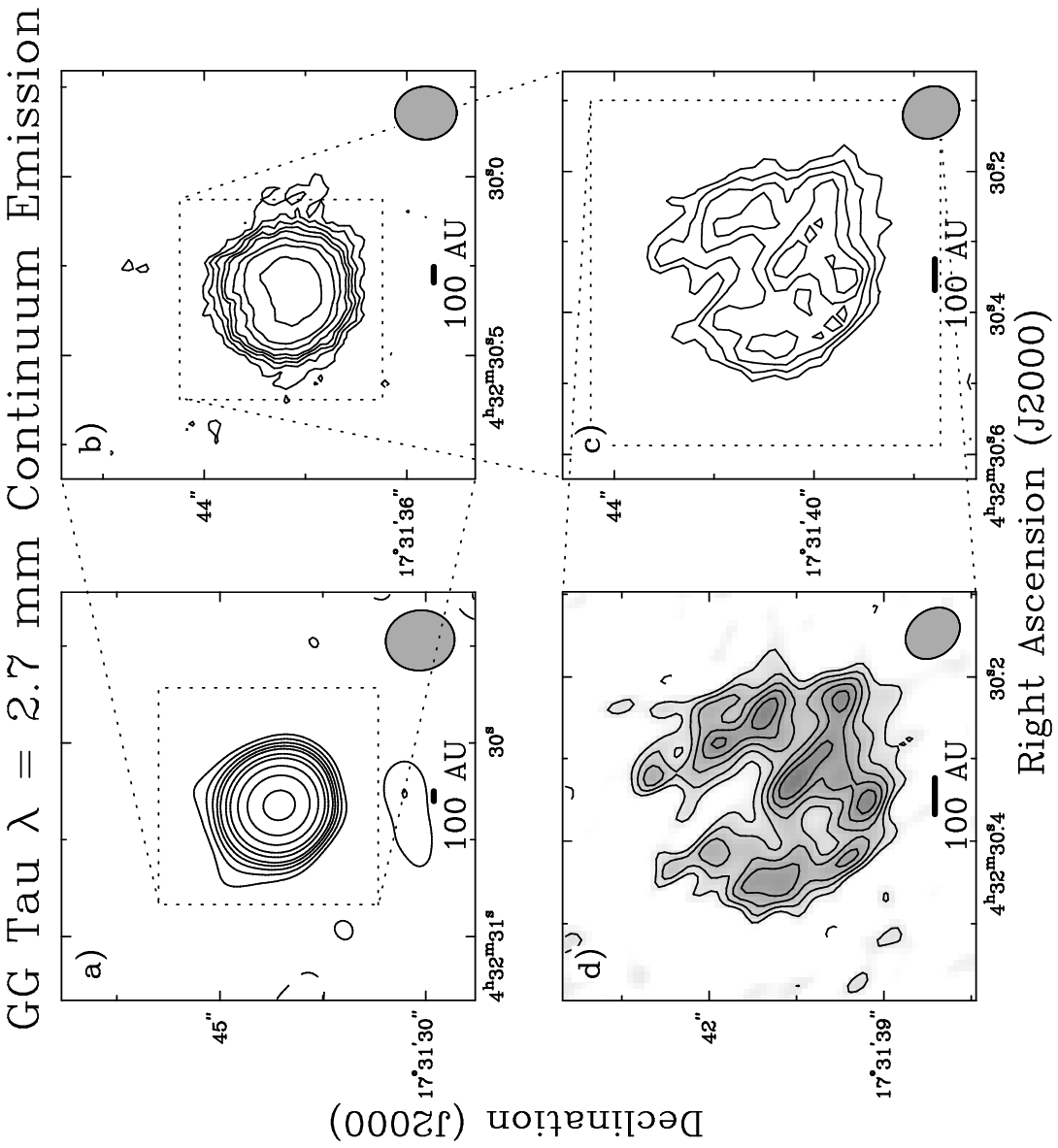}
\vspace{4.5in}
\caption{
GG Tauri maps of the $\lambda$ = 2.7 mm continuum emission.
Panel (a) is contoured in steps of (-4 -3 -2 2 3 4 5 6 8 10 14.14 20 28.28)
$\times$ the rms of panel (a) of 1.8 mJy/beam.
Panels (b) through (d) are (-4 -3 -2 2 3 4 5 6 8 10 14.14)
$\times$ a rms noise of 1.5 mJy/beam.
(a) $\sigma$ = 1.8 mJy/beam; beam is 5$\farcs$02 $\times$ 4$\farcs$44 P.A. =   7$\arcdeg$.
(b) $\sigma$ = 1.2 mJy/beam; beam is 2$\farcs$44 $\times$ 2$\farcs$12 P.A. =   1$\arcdeg$.
(c) $\sigma$ = 1.2 mJy/beam; beam is 1$\farcs$17 $\times$ 1$\farcs$02 P.A. =  31$\arcdeg$.
(d) $\sigma$ = 1.5 mJy/beam; beam is 1$\farcs$02 $\times$ 0$\farcs$80 P.A. =  40$\arcdeg$.
The greyscale is used to emphasize the hills and valleys of the ``clumps'' in the circumbinary
disk.}
\end{figure}
\clearpage

\begin{figure}
\includegraphics{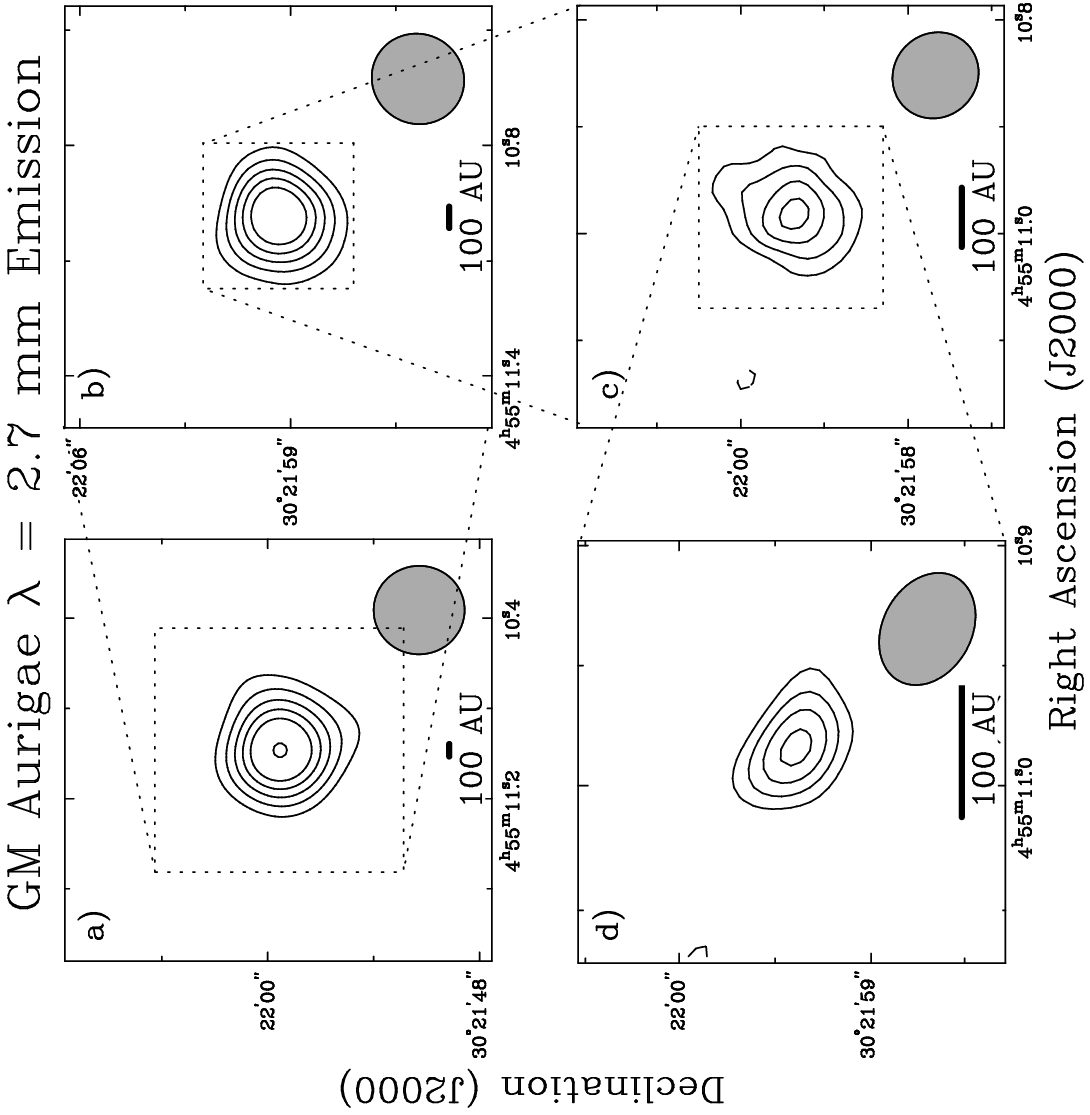}
\vspace{4.5in}
\caption{
GM Aurigae maps of the $\lambda$ = 2.7 mm continuum emission.
All panels are contoured in steps of (-4 -3 -2 2 3 4 5 6 8 10 14.14 20 28.28)  
$\times$ a rms noise of 2.5 mJy/beam.
(a) $\sigma$ = 1.1 mJy/beam; beam is 5$\farcs$15 $\times$ 5$\farcs$08 P.A. =   6$\arcdeg$.
(b) $\sigma$ = 0.9 mJy/beam; beam is 3$\farcs$12 $\times$ 3$\farcs$00 P.A. = -30$\arcdeg$.
(c) $\sigma$ = 1.6 mJy/beam; beam is 1$\farcs$07 $\times$ 1$\farcs$00 P.A. =  51$\arcdeg$.
(d) $\sigma$ = 2.5 mJy/beam; beam is 0$\farcs$63 $\times$ 0$\farcs$47 P.A. =  62$\arcdeg$.}
\end{figure}
\clearpage

\begin{figure}
\includegraphics{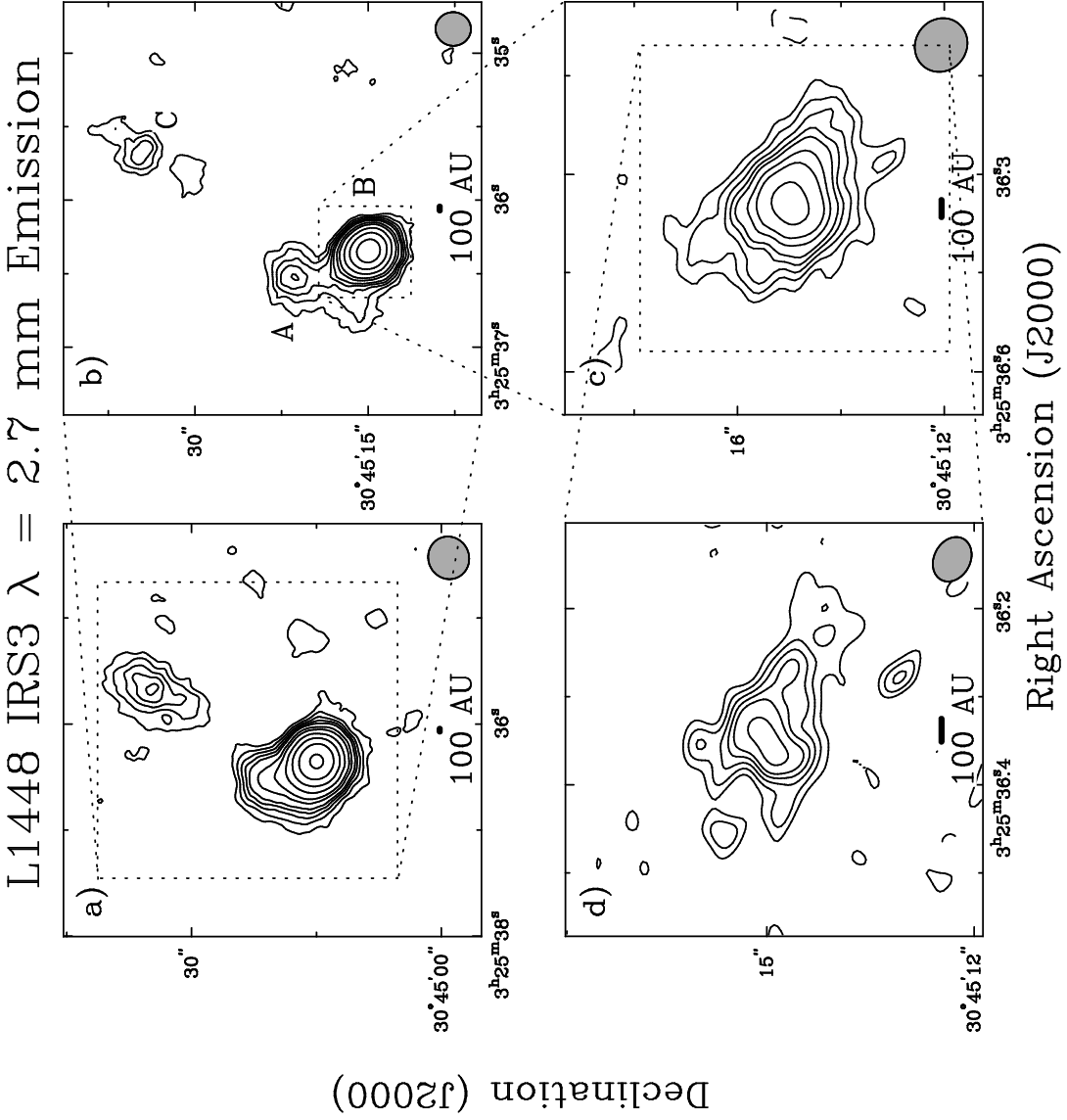}
\vspace{4.5in}
\caption{
L1448 IRS3 maps of the $\lambda$ = 2.7 mm continuum emission.
All panels are contoured in steps of (-4 -3 -2 2 3 4 5 6 8 10 14.14 20 28.28 40 56.56)
$\times$ a rms noise of 2.3 mJy/beam.
(a) $\sigma$ = 1.6 mJy/beam; beam is 5$\farcs$22 $\times$ 4$\farcs$89 P.A. = -71$\arcdeg$.
(b) $\sigma$ = 1.5 mJy/beam; beam is 3$\farcs$06 $\times$ 2$\farcs$95 P.A. = -61$\arcdeg$.
(c) $\sigma$ = 1.6 mJy/beam; beam is 1$\farcs$08 $\times$ 0$\farcs$99 P.A. =  56$\arcdeg$.
(d) $\sigma$ = 2.3 mJy/beam; beam is 0$\farcs$68 $\times$ 0$\farcs$52 P.A. =  63$\arcdeg$.}
\end{figure}
\clearpage

\begin{figure}
\includegraphics{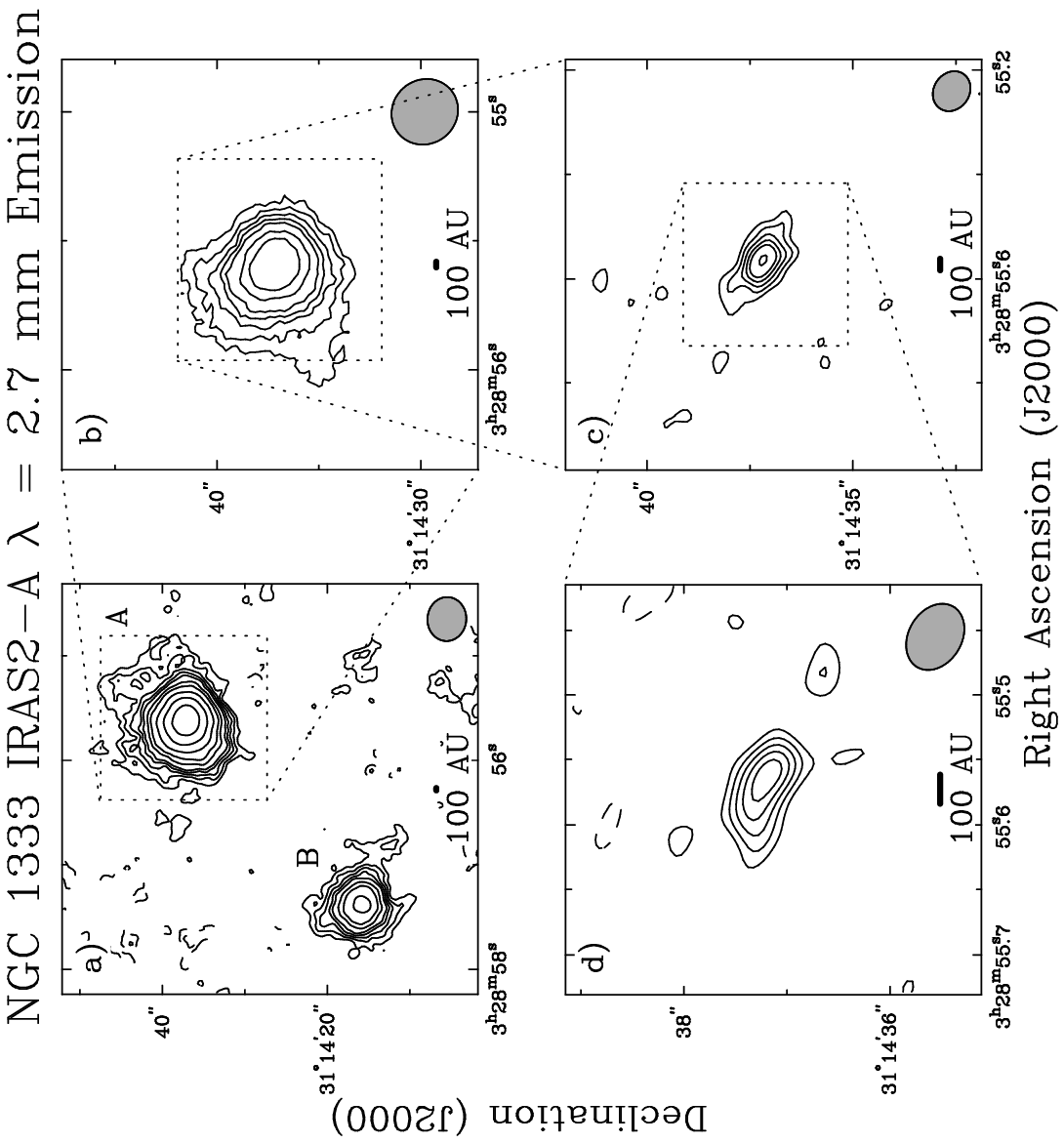}
\vspace{4.5in}
\caption{
NGC 1333 IRAS2 A maps of the $\lambda$ = 2.7 mm continuum emission.
Panel (a) is contoured in steps of (-4 -3 -2 2 3 4 5 6 8 10 14.14 20 28.28 40 56.56)
$\times$ the rms of panel (a) of 1.3 mJy/beam.
Panels (b) through (d) are (-4 -3 -2 2 3 4 5 6 8 10 14.14 20 28.28)
$\times$ a rms noise of 2.7 mJy/beam.
(a) $\sigma$ = 1.3 mJy/beam; beam is 5$\farcs$40 $\times$ 4$\farcs$70 P.A. =  86$\arcdeg$.
(b) $\sigma$ = 1.2 mJy/beam; beam is 3$\farcs$36 $\times$ 3$\farcs$16 P.A. =  45$\arcdeg$.
(c) $\sigma$ = 1.7 mJy/beam; beam is 1$\farcs$02 $\times$ 0$\farcs$87 P.A. =  57$\arcdeg$.
(d) $\sigma$ = 2.7 mJy/beam; beam is 0$\farcs$69 $\times$ 0$\farcs$52 P.A. =  60$\arcdeg$.}
\end{figure}
\clearpage

\begin{figure}
\includegraphics{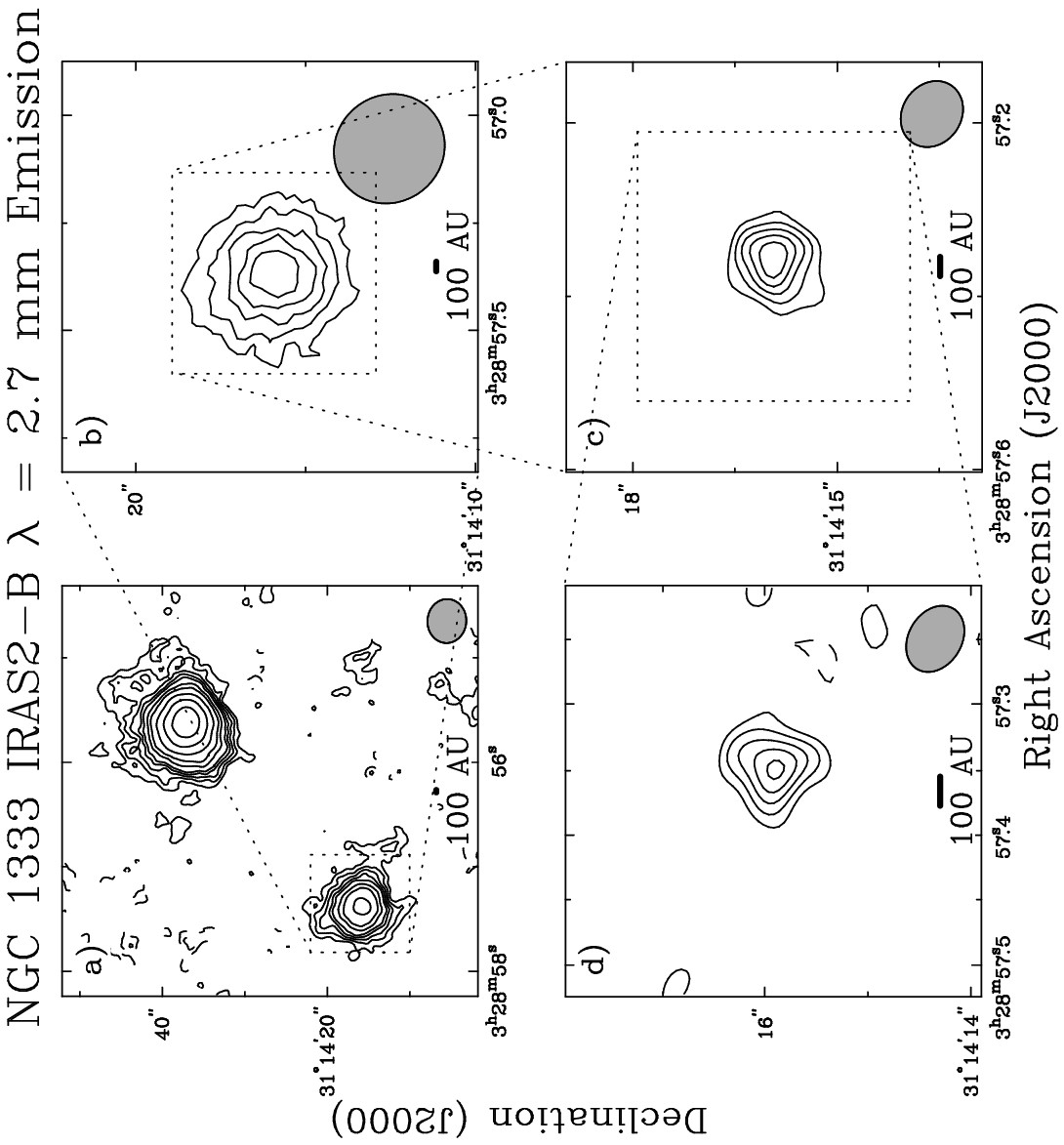}
\vspace{4.5in}
\caption{
NGC 1333 IRAS2 B maps of the $\lambda$ = 2.7 mm continuum emission.
Panel (a) is contoured in steps of (-4 -3 -2 2 3 4 5 6 8 10 14.14 20 28.28 40 56.56)
$\times$ the rms of panel (a) of 1.3 mJy/beam.
Panels (b) through (d) are (-4 -3 -2 2 3 4 5 6 8 10 14.14)
$\times$ a rms noise of 2.7 mJy/beam.
(a) $\sigma$ = 1.3 mJy/beam; beam is 5$\farcs$40 $\times$ 4$\farcs$70 P.A. =  86$\arcdeg$.
(b) $\sigma$ = 1.2 mJy/beam; beam is 3$\farcs$36 $\times$ 3$\farcs$16 P.A. =  45$\arcdeg$.
(c) $\sigma$ = 1.7 mJy/beam; beam is 1$\farcs$02 $\times$ 0$\farcs$87 P.A. =  57$\arcdeg$.
(d) $\sigma$ = 2.7 mJy/beam; beam is 0$\farcs$69 $\times$ 0$\farcs$52 P.A. =  60$\arcdeg$.}
\end{figure}
\clearpage

\begin{figure}
\includegraphics{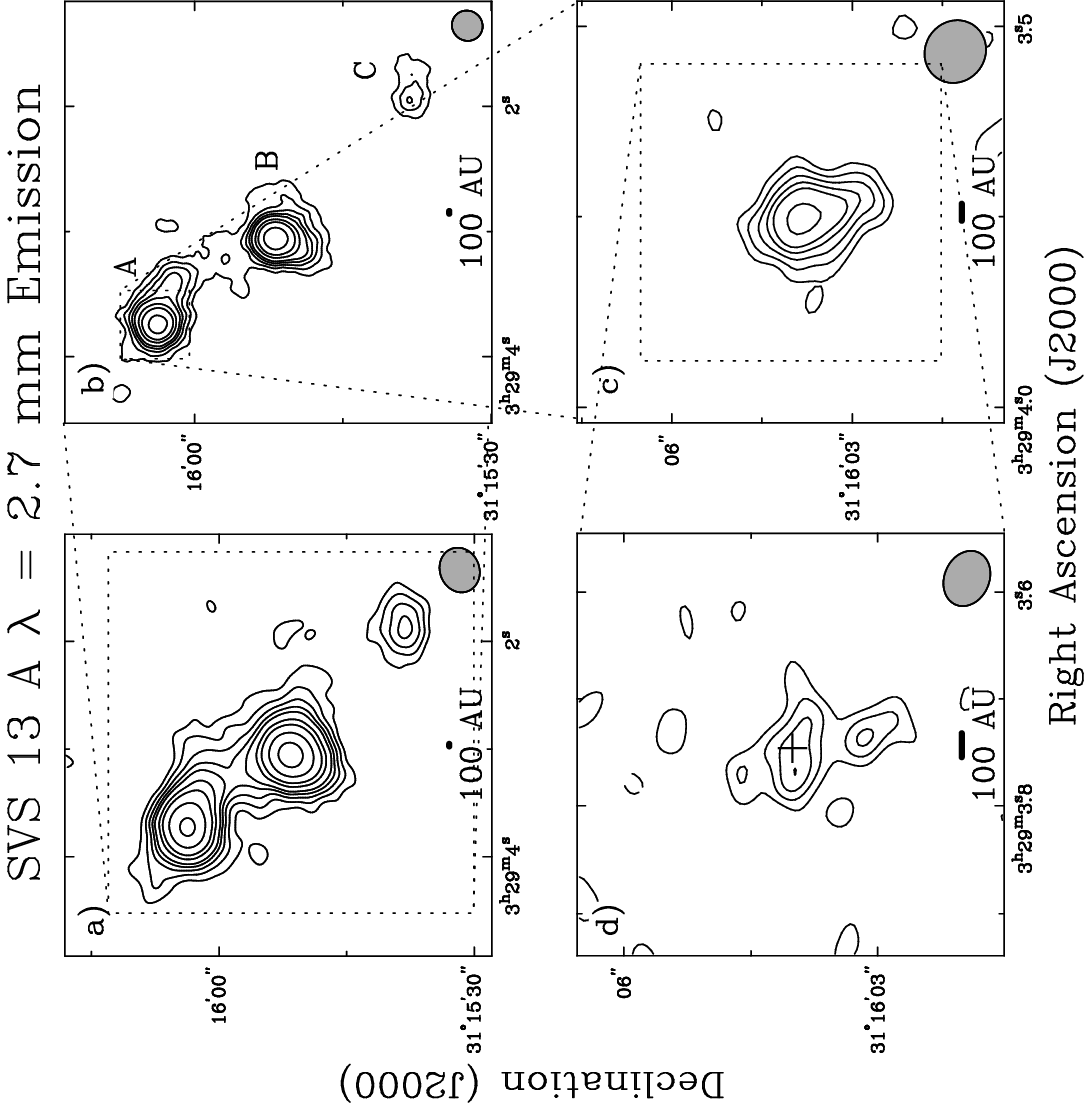}
\vspace{4.5in}
\caption{
SVS13 A maps of the $\lambda$ = 2.7 mm continuum emission.
All panels are contoured in steps of (-4 -3 -2 2 3 4 5 6 8 10 14.14 20 28.28 40)
$\times$ a rms noise of 2.2 mJy/beam.
(a) $\sigma$ = 1.1 mJy/beam; beam is 5$\farcs$40 $\times$ 4$\farcs$64 P.A. = -70$\arcdeg$.
(b) $\sigma$ = 1.1 mJy/beam; beam is 3$\farcs$17 $\times$ 3$\farcs$05 P.A. = -43$\arcdeg$.
(c) $\sigma$ = 1.5 mJy/beam; beam is 1$\farcs$08 $\times$ 1$\farcs$00 P.A. =  57$\arcdeg$.
(d) $\sigma$ = 2.2 mJy/beam; beam is 0$\farcs$68 $\times$ 0$\farcs$53 P.A. =  68$\arcdeg$.
The cross in panel (d) is the $\lambda$ = 3.6 cm peak from Rodr\'{i}guez, Anglada, \& Curiel (1997).}
\end{figure}
\clearpage

\begin{figure}
\includegraphics{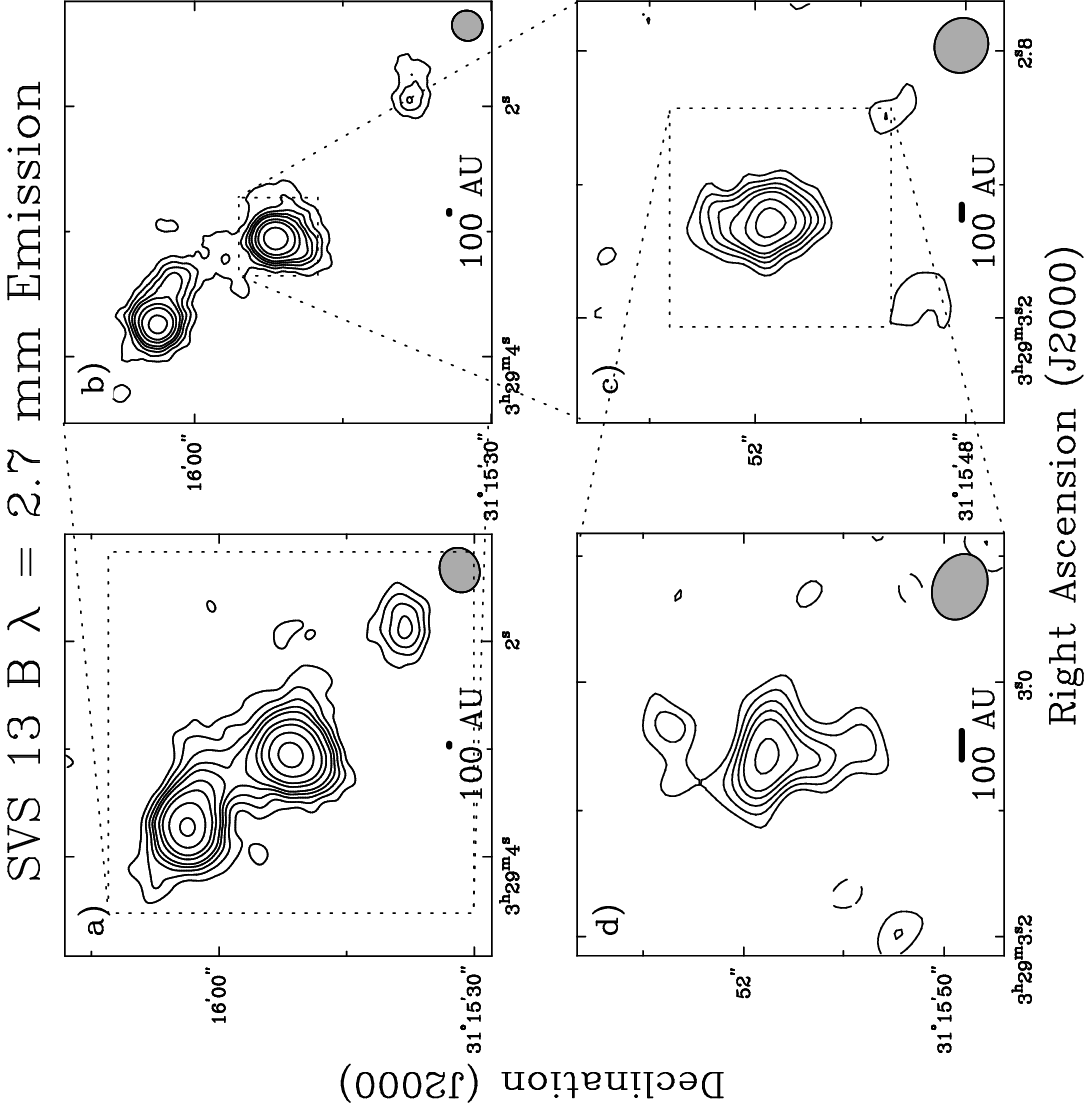}
\vspace{4.5in}
\caption{
SVS13 B maps of the $\lambda$ = 2.7 mm continuum emission.
All panels are contoured in steps of (-4 -3 -2 2 3 4 5 6 8 10 14.14 20 28.28 40)
$\times$ a rms noise of 2.2 mJy/beam.
(a) $\sigma$ = 1.1 mJy/beam; beam is 5$\farcs$40 $\times$ 4$\farcs$64 P.A. = -70$\arcdeg$.
(b) $\sigma$ = 1.1 mJy/beam; beam is 3$\farcs$17 $\times$ 3$\farcs$05 P.A. = -43$\arcdeg$.
(c) $\sigma$ = 1.5 mJy/beam; beam is 1$\farcs$08 $\times$ 1$\farcs$00 P.A. =  57$\arcdeg$.
(d) $\sigma$ = 2.2 mJy/beam; beam is 0$\farcs$68 $\times$ 0$\farcs$53 P.A. =  68$\arcdeg$.}
\end{figure}
\clearpage

\begin{figure}
\includegraphics{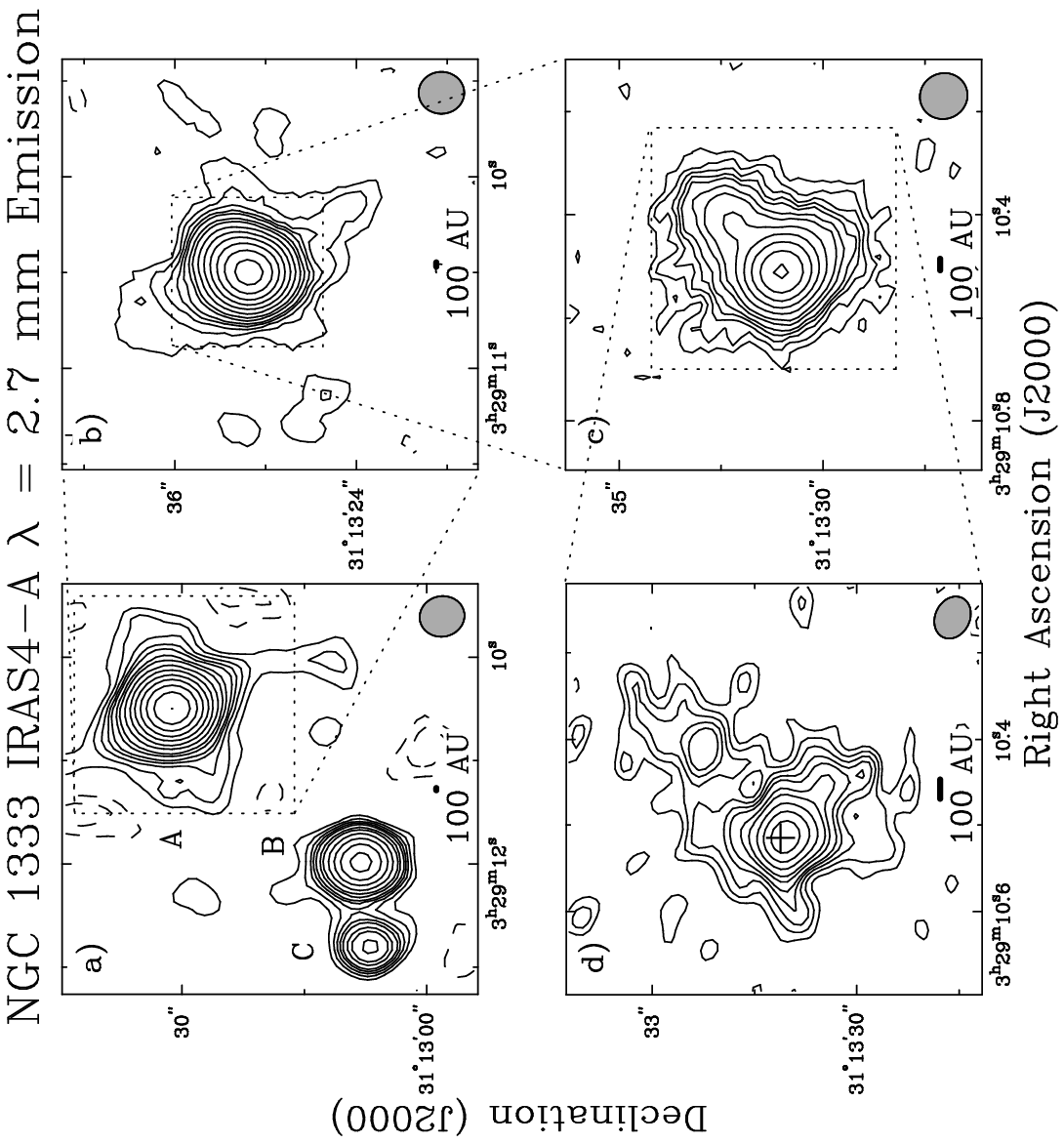}
\vspace{4.5in}
\caption{
NGC 1333 IRAS4 A maps of the $\lambda$ = 2.7 mm continuum emission.
Panel (a) is contoured in steps of (-4 -3 -2 2 3 4 5 6 8 10 14.14 20 28.28 40)
$\times$ the rms of panel (a) of 3.1 mJy/beam.
Panels (b) through (d) are (-4 -3 -2 2 3 4 5 6 8 10 14.14 20 28.28 40 56.56)
$\times$ a rms noise of 2.9 mJy/beam.
(a) $\sigma$ = 3.1 mJy/beam; beam is 5$\farcs$52 $\times$ 5$\farcs$02 P.A. =  12$\arcdeg$.
(b) $\sigma$ = 1.9 mJy/beam; beam is 3$\farcs$02 $\times$ 2$\farcs$81 P.A. =   1$\arcdeg$.
(c) $\sigma$ = 2.1 mJy/beam; beam is 1$\farcs$18 $\times$ 1$\farcs$13 P.A. =  30$\arcdeg$.
(d) $\sigma$ = 2.9 mJy/beam; beam is 0$\farcs$65 $\times$ 0$\farcs$51 P.A. =  65$\arcdeg$.
The cross in panel (d) is the $\lambda$ = 1.3 cm peak from Mundy et al. (1993).}
\end{figure}
\clearpage

\begin{figure}
\includegraphics{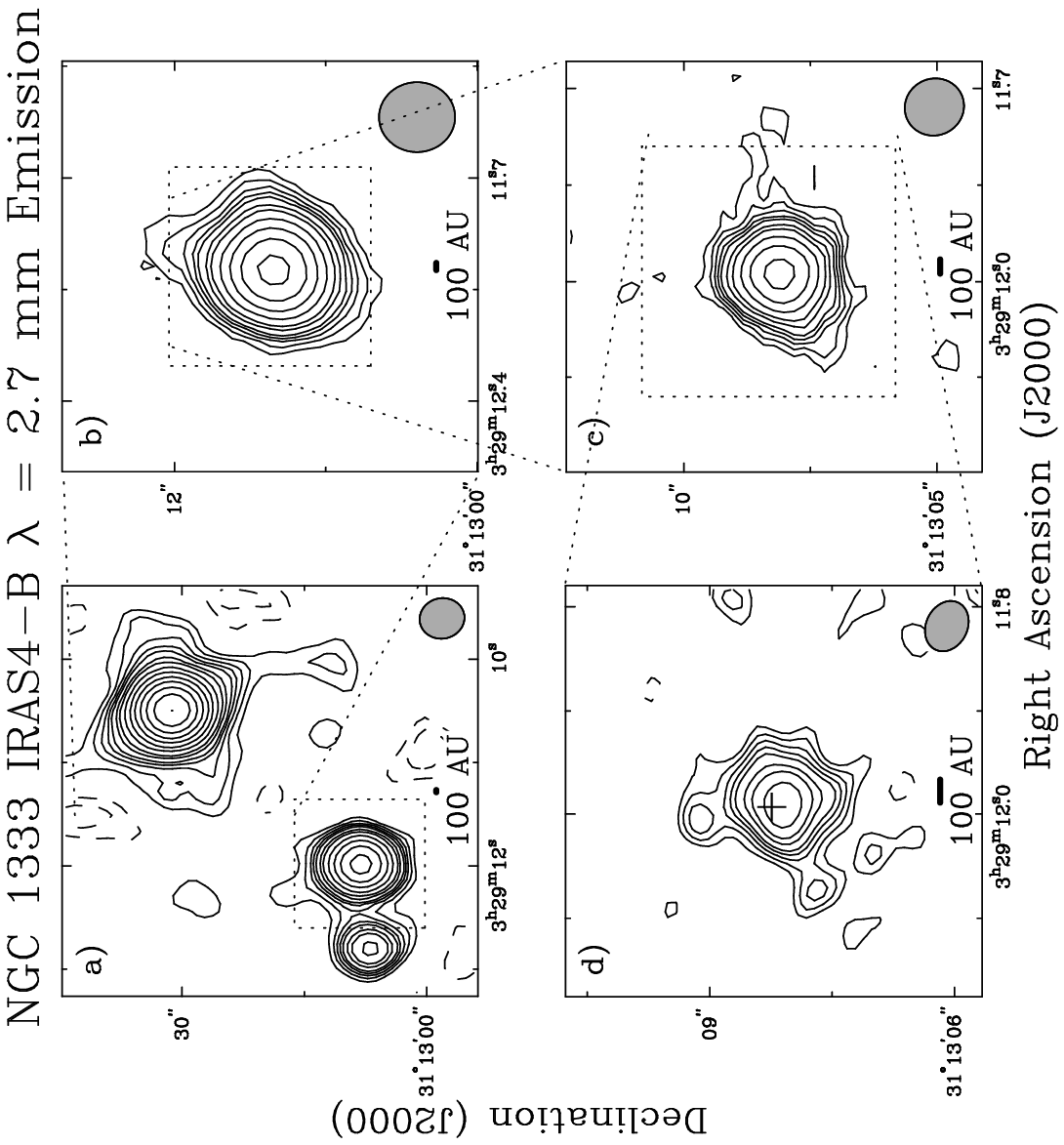}
\vspace{4.5in}
\caption{
NGC 1333 IRAS4 B maps of the $\lambda$ = 2.7 mm continuum emission.
Panel (a) is contoured in steps of (-4 -3 -2 2 3 4 5 6 8 10 14.14 20 28.28 40)
$\times$ the rms of panel (a) of 3.1 mJy/beam.
Panels (b) through (d) are (-4 -3 -2 2 3 4 5 6 8 10 14.14 20 28.28 40 56.56)
$\times$ a rms noise of 2.9 mJy/beam.
(a) $\sigma$ = 3.1 mJy/beam; beam is 5$\farcs$52 $\times$ 5$\farcs$02 P.A. =  12$\arcdeg$.
(b) $\sigma$ = 1.9 mJy/beam; beam is 3$\farcs$02 $\times$ 2$\farcs$81 P.A. =   1$\arcdeg$.
(c) $\sigma$ = 2.1 mJy/beam; beam is 1$\farcs$18 $\times$ 1$\farcs$13 P.A. =  30$\arcdeg$.
(d) $\sigma$ = 2.9 mJy/beam; beam is 0$\farcs$65 $\times$ 0$\farcs$51 P.A. =  65$\arcdeg$.
The cross in panel (d) is the $\lambda$ = 1.3 cm peak from Mundy et al. (1993).}
\end{figure}
\clearpage

\begin{figure}
\includegraphics{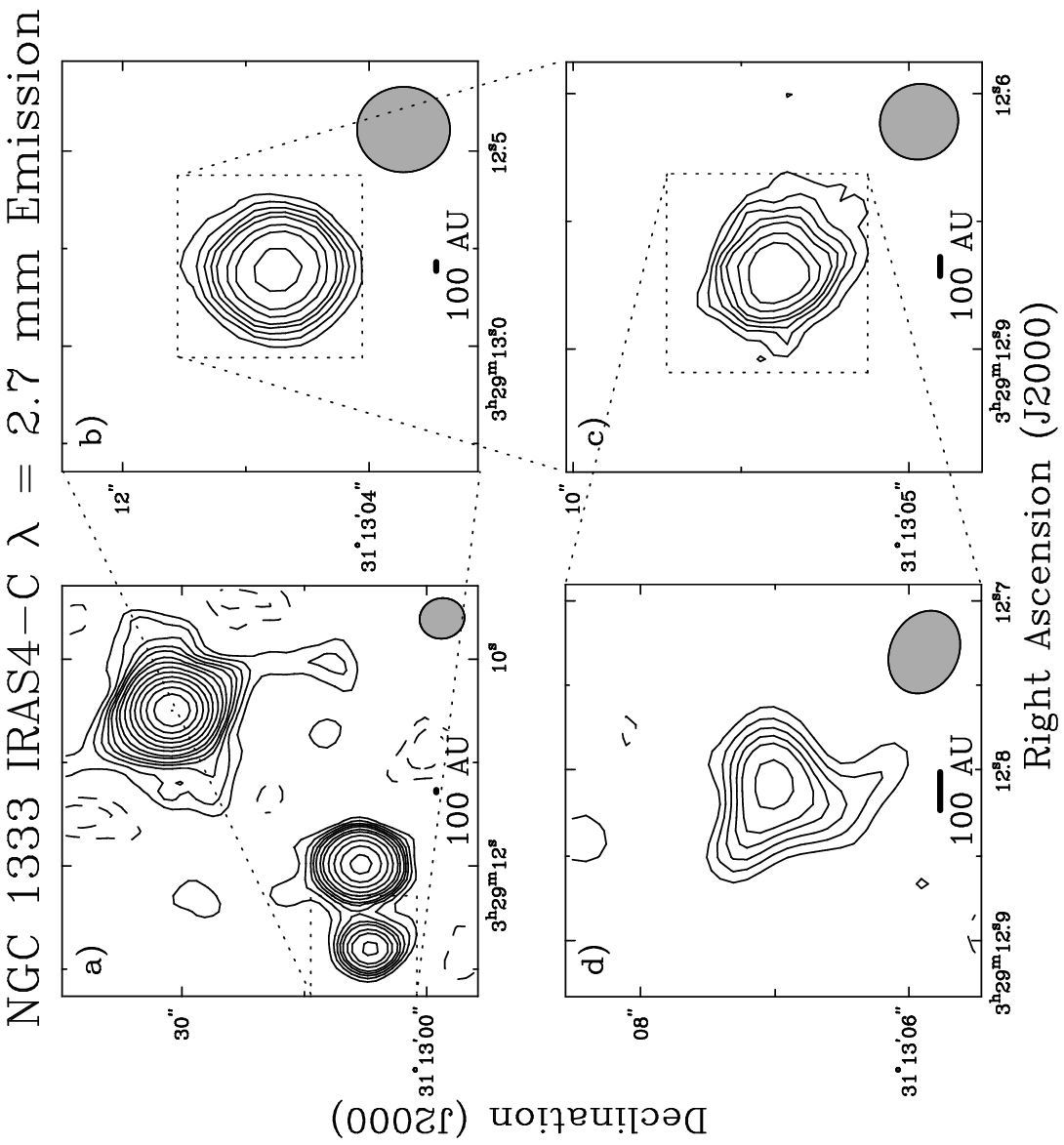}
\vspace{4.5in}
\caption{
NGC 1333 IRAS4 C maps of the $\lambda$ = 2.7 mm continuum emission.
Panel (a) is contoured in steps of (-4 -3 -2 2 3 4 5 6 8 10 14.14 20 28.28 40)
$\times$ the rms of panel (a) of 3.1 mJy/beam.
Panels (b) through (d) are (-4 -3 -2 2 3 4 5 6 8 10 14.14 20 28.28 40 56.56)
$\times$ a rms noise of 2.9 mJy/beam.
(a) $\sigma$ = 3.1 mJy/beam; beam is 5$\farcs$52 $\times$ 5$\farcs$02 P.A. =  12$\arcdeg$.
(b) $\sigma$ = 1.9 mJy/beam; beam is 3$\farcs$02 $\times$ 2$\farcs$81 P.A. =   1$\arcdeg$.
(c) $\sigma$ = 2.1 mJy/beam; beam is 1$\farcs$18 $\times$ 1$\farcs$13 P.A. =  30$\arcdeg$.
(d) $\sigma$ = 2.9 mJy/beam; beam is 0$\farcs$65 $\times$ 0$\farcs$51 P.A. =  65$\arcdeg$.}
\end{figure}
\clearpage

\begin{figure}
\includegraphics{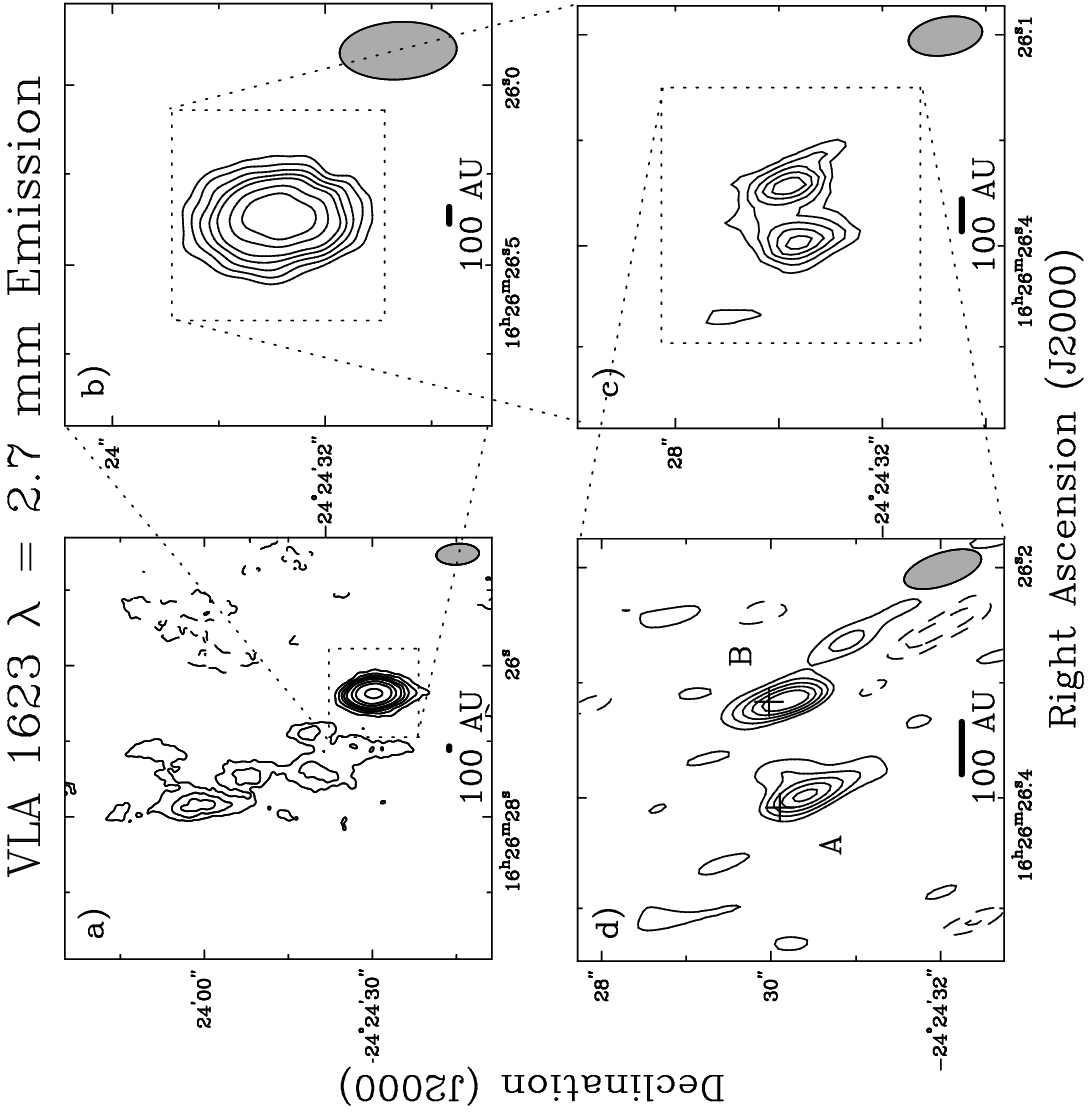}
\vspace{4.5in}
\caption{
VLA 1623 maps of the $\lambda$ = 2.7 mm continuum emission.
All panels are contoured in steps of (-4 -3 -2 2 3 4 5 6 8 10 14.14)  
$\times$ a rms noise of 3.5 mJy/beam.
(a) $\sigma$ = 3.0 mJy/beam; beam is 7$\farcs$65 $\times$ 3$\farcs$80 P.A. =   4$\arcdeg$.
(b) $\sigma$ = 2.2 mJy/beam; beam is 4$\farcs$40 $\times$ 2$\farcs$19 P.A. =   3$\arcdeg$.
(c) $\sigma$ = 2.0 mJy/beam; beam is 1$\farcs$44 $\times$ 0$\farcs$74 P.A. =  10$\arcdeg$.
(d) $\sigma$ = 3.5 mJy/beam; beam is 0$\farcs$95 $\times$ 0$\farcs$39 P.A. =  18$\arcdeg$.
The crosses indicate the $\lambda$ = 3.6 cm positions from 
Bontemps \& Andr\`{e} (1997).}
\end{figure}
\clearpage

\begin{figure}
\includegraphics{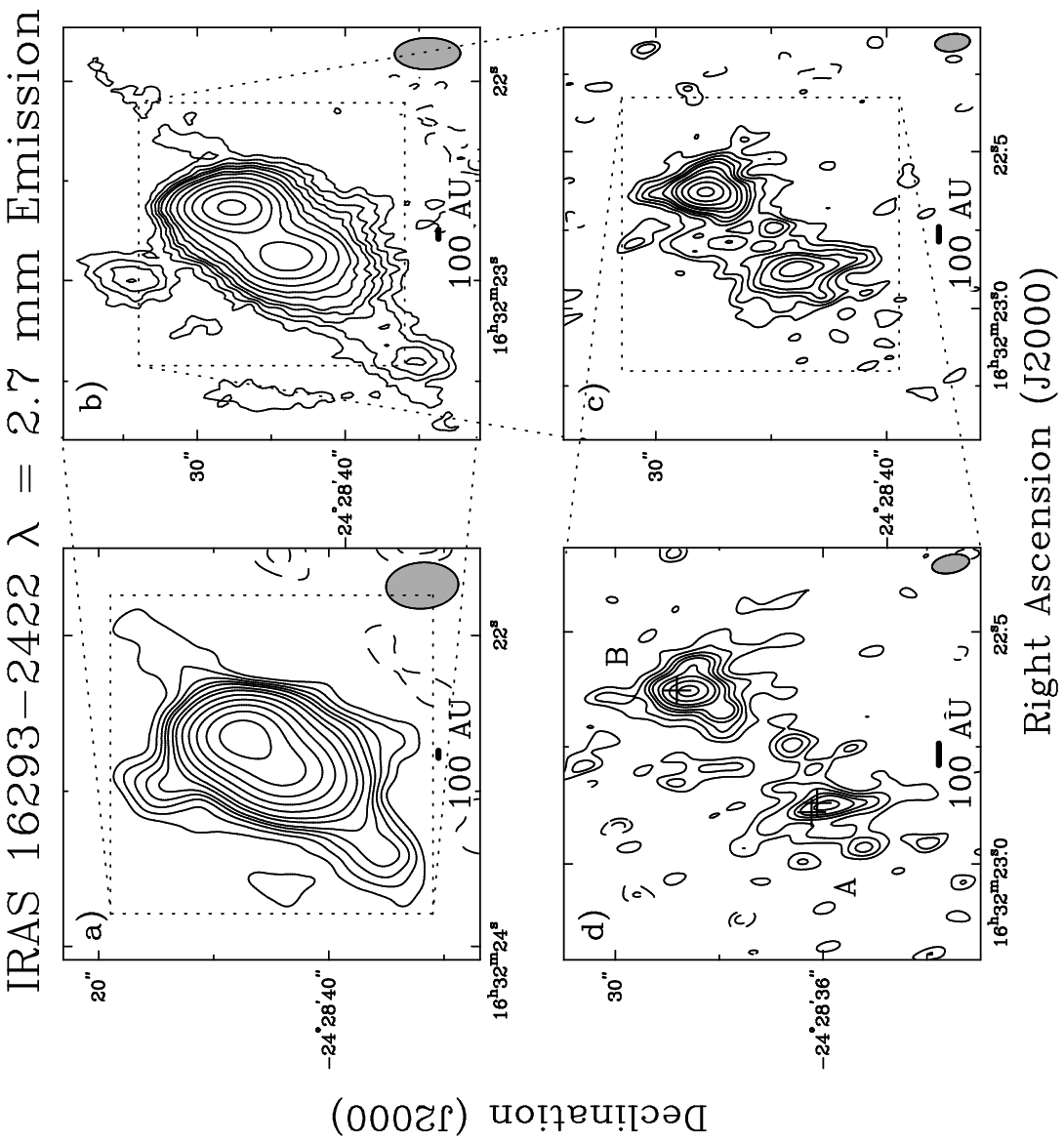}
\vspace{4.5in}
\caption{
IRAS 16293-2422 maps of the $\lambda$ = 2.7 mm continuum emission.
Panel (a) is contoured in steps of (-4 -3 -2 2 3 4 5 6 8 10 14.14 20 28.28 40 56.56)
$\times$ the rms of panel (a) of 5.8 mJy/beam.
Panels (b) through (d) are (-4 -3 -2 2 3 4 5 6 8 10 14.14 20 28.28 40 56.56)
$\times$ a rms noise of 4.8 mJy/beam.
(a) $\sigma$ = 5.8 mJy/beam; beam is 6$\farcs$29 $\times$ 4$\farcs$06 P.A. =   4$\arcdeg$.
(b) $\sigma$ = 4.2 mJy/beam; beam is 4$\farcs$45 $\times$ 2$\farcs$16 P.A. =   1$\arcdeg$.
(c) $\sigma$ = 4.1 mJy/beam; beam is 1$\farcs$52 $\times$ 0$\farcs$76 P.A. =   7$\arcdeg$.
(d) $\sigma$ = 4.8 mJy/beam; beam is 1$\farcs$09 $\times$ 0$\farcs$53 P.A. =  11$\arcdeg$.
The crosses indicate the $\lambda$ = 2 cm positions from Wootten (1989).}
\end{figure}
\clearpage

\begin{figure}
\includegraphics{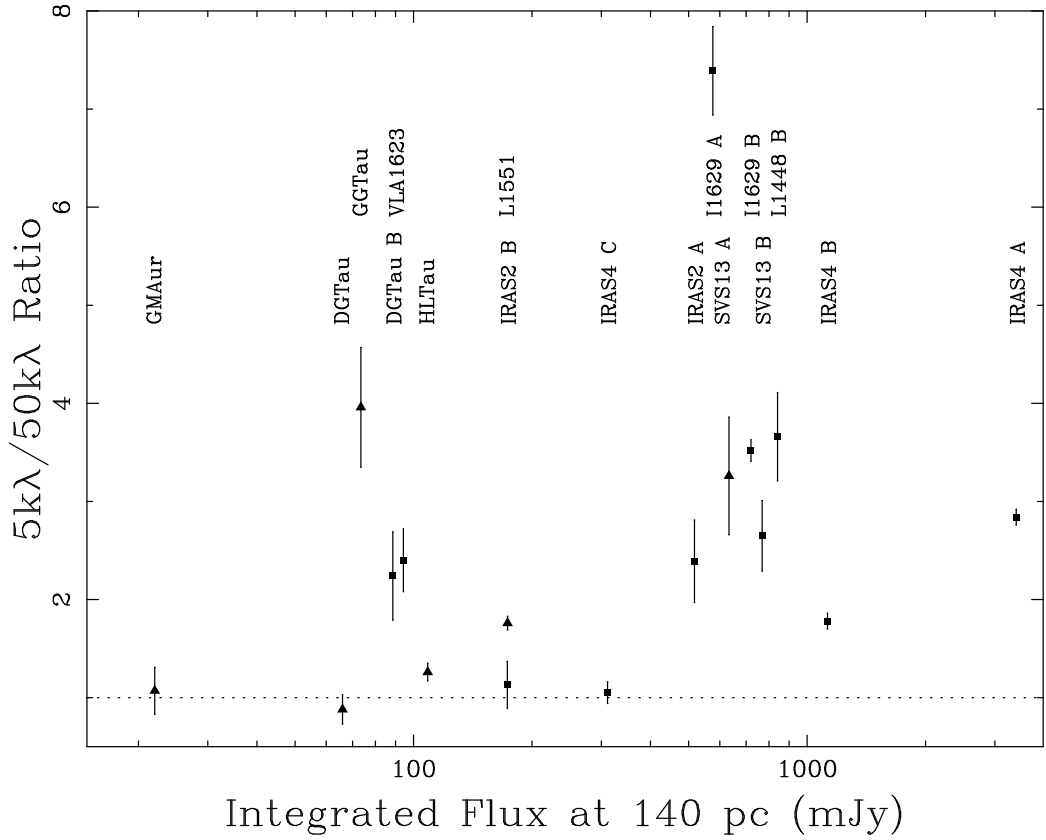}
\vspace{4.5in}
\caption{
Comparison of the ratio of the flux at 5k$\lambda$ and 50k$\lambda$ fringe spacings
amplitude and the integrated flux of each object from Table 2, both adjusted to the
distance of Taurus.
The solid triangle symbols indicate optical/IR sources and the solid square symbols indicate the
embedded sources.  
Each point is labeled with its corresponding source.}
\end{figure}
\clearpage

\begin{figure}
\includegraphics{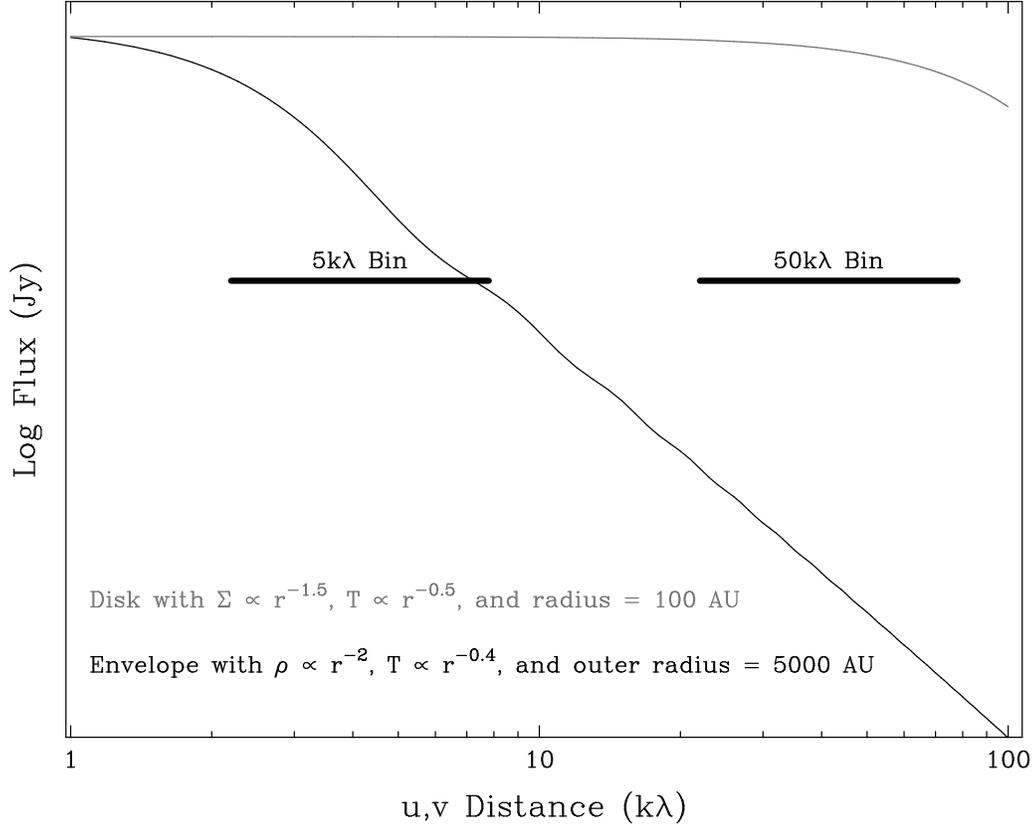}
\vspace{4.5in}
\caption{
Model visibility curves of a 100 AU radius disk and a 5000 AU outer radius envelope
at the distance of Taurus.
These simple models assume radial power-law for both temperature and density
(volume density, $\rho$, for the envelope model and surface density,
$\Sigma$, for the disk model).
The horizontal axis is antenna separation in kilo-wavelengths, and the vertical axis
is arbitrary flux in Jy.
A disk will have a 5k$\lambda$/50k$\lambda$ 
ratio $\sim$~1, and an envelope with an outer radius of 
$\ge$ 1000 AU will have a 5k$\lambda$/50k$\lambda$ ratio $>$~1.
}
\end{figure}


\begin{references}
\reference{a90} Adams, F.C., Emerson, J.P., \& Fuller, G.A. 1990, \apj, 357, 606
\reference{a87} Adams, F.C., Lada, C.J., \& Shu, F.H. 1987, \apj, 312, 788
\reference{a89} Adams, F.C., Ruden, S.P., \& Shu, F.H. 1989, \apj, 347, 959
\reference{an90} Andr\'{e}, P., Mart\'{i}n-Pintado, J., Despois, D., \& Montmerle, T. 1990, \aap, 236, 180
\reference{an93} Andr\'{e}, P., Ward-Thompson, D., \& Barsony, M. 1993, \apj, 406, 122
\reference{ang} Anglada, G., Rodr\'{i}guez, L.F., Torrelles, J.M., Estalella, R., Ho, P.T.P, Cant\'{o}, J., L\'{o}pez, R., \& Verdes-Montenegro, L. 1989, \apj, 341, 208
\reference{art} Artymowiez, P., \& Lubow, S.H. 1994, \apj, 421, 651
\reference{asp} Aspin, C., Sandell, G., \& Russell, A.P.G. 1994, \aaps, 106, 165
\reference{ba91} Bachiller, R., Andr\'{e}, P., \& Cabrit, S. 1991, \aap, 241, L43
\reference{ba86} Bachiller, R., \& Cernicharo, J. 1986, \aap, 166, 283
\reference{ba90} Bachiller, R., Cernicharo, J., Mart\'{i}n-Pintado, J., Tafalla, M., \& Lazareff, B. 1990, \aap, 231, 174
\reference{ba95} Bachiller, R., Guilloteau, S., Dutrey, A., Planesas, P., \& Mart\'{i}n-Pintado, J. 1995, \aap, 299, 857
\reference{ba98} Bachiller, R., Guilloteau, S., Gueth, F., Tafalla, M., Dutrey, A., Codella, C., \& Castets, A. 1998, \aap, 339, L49 
\reference{ball} Bally, J., Devine, D., \& Reipurth, B. 1996, \apj, 473, L49
\reference{bar} Barsony, M., Ward-Thompson, D., Andr\'{e}, P., \& O'Linger, J. 1998, \apj, 509, 733
\reference{basu} Basu, S. 1998, \apj, 509, 229
\reference{bat97} Bate, M.R., \& Bonnell, I.A. 1997, \mnras, 285, 33
\reference{bat} Bate, M.R., Bonnell, I.A., \& Price, N.M. 1995, \mnras, 277, 362
\reference{b95} Beckwith, S.V.W., \& Birk, C.C. 1995, \apj, 449, L59
\reference{b91} Beckwith, S.V.W., \& Sargent, A.I. 1991, \apj, 381, 250
\reference{b90} Beckwith, S.V.W., Sargent, A.I., Chini, R.S., \& G\"{u}sten, R. 1990, \aj, 99, 924
\reference{bert} Bertout, C., Basri, G., \& Bouvier, J. 1988, \apj, 330, 350
\reference{bie} Bieging, J., Cohen, M., \& Schwartz, P.R. 1984, \apj, 282, 699
\reference{bla} Blake, G.A. 1997, IAU Symposium 178, Molecules in Astrophysics: Probes and Processes, ed. E.F. van Dishoeck, (Dordrecht: Kluwer), 31
\reference{bon94} Bonnell, I.A. 1994, \mnras, 269, 837
\reference{bon92} Bonnell, I.A., \& Bastein, P. 1992, \apj, 401, 654
\reference{bon94b} Bonnell, I.A., \& Bate, M.R. 1994, \mnras, 271, 999 
\reference{bon97} Bonnell, I.A., Bate, M.R., Clarke, C.J., \& Pringle, J.E. 1997, \mnras, 285, 201
\reference{bon91} Bonnell, I.A., Martel, H., Bastein, P., Arcoragi, J.P., \& Benz, W. 1991, \apj, 377, 553
\reference{bon} Bontemps, S., \&  Andr\'{e}, P. 1997, Low Mass Star Formation from Infall to Outflow, Poster proceedings of IAU Symposium 182, Low Herbig-Haro Flows and the Birth of Low Mass Stars, ed. F. Malbet \& A. Castets, 63
\reference{bo93} Boss, A.P. 1993, \apj, 410, 157
\reference{bo95} Boss, A.P. 1995, Rev. Mex. A.A., 1, 165
\reference{bo97} Boss, A.P. 1997, \apj, 483, 309
\reference{bo79} Boss, A.P., \& Bodenheimer, P. 1979, \apj, 234, 289
\reference{bov} Bouvier, J., Rigaut, F., \& Nadeau, D. 1997, \aap, 323, 139
\reference{brig} Briggs, D.S. 1995, BAAS, 187, 112.02
\reference{burk} Burkert, A., \& Bodenheimer, P. 1993, \mnras, 264, 798
\reference{cal} Calvet, N, Hartmann, L., Kenyon, S.J., \& Whitney, B.A. 1994, \apj, 434, 330
\reference{cas} Cassen, P., \& Moosman, A. 1981, Icarus, 48, 353
\reference{cer} Cernis, K. 1983, \apss, 166, 315
\reference{chini} Chini, R., Reipurth, B., Sievers, A., Ward-Thompson, D., Haslam, C.G.T., Kreysa, E., \& Lemke, R. 1997, \aap, 325, 542
\reference{close} Close, L.M., Roddier, F., Northcott, M.J., Roddier, C., \& Graves, J.E. 1995, \apj, 478, 766
\reference{cb86} Cohen, M., \& Bieging, J.H. 1986, \aj, 92, 6
\reference{cur} Curiel, S., Raymond, J.C., Rodr\'{i}guez, L.F., Cant\'{o}, J., \& Moran, J.M. 1990, \apj, 365, L85
\reference{dav} Davis, C.J., \& Smith, M.D. 1995, \apj, 443, L41
\reference{den} Dent, W.R.F., Matthews, H.E., \& Walther, D.M. 1995, \mnras, 277, 193
\reference{dra} Draine, B.T. 1990, The Interstellar Medium in Galaxies, ed. H.A. Thronson \& J.M. Shull, (Dordrecht: Kluwer), 483 
\reference{dur} Durisen, R.H., \& Sterzik, M.F. 1994, \aap, 286, 84
\reference{d96} Dutrey, A., Guilloteau, S., Duvert, G., Prato, L., Simon, M., Schuster, K., \& M\'{e}nard, F. 1996, \aap, 309, 493
\reference{d94} Dutrey, A., Guilloteau, S., \& Simon, M. 1994, \aap, 286, 149
\reference{duq} Duquennoy, A., \& Mayor, M. 1991, \aap, 248, 485
\reference{eis} Eisl\"{u}ffel, J., G\"{u}nther, E., Hessman, F.V., Mundt, R., Poetzel, R., Carr, J.S., Beckwith, S.V.W., \& Ray, T.P. 1991, \apj, 383, L19
\reference{elias} Elias, J. 1978, \apj, 224, 857
\reference{galli} Galli, D., \& Shu, F.H. \apj, 417, 220
\reference{gh93} Ghez, A.M., Neugebauer, G., Matthews, K. 1993, \aj, 106, 2005
\reference{gh97} Ghez, A.M., White, R.J., \& Simon, M. 1997, \apj, 490, 353
\reference{gross} Grossman, E.N., Masson, C.R., Sargent, A>I., Scoville, N.Z., Scott, S., \& Woody, D.P. 1987, \apj, 320, 356
\reference{hay} Hayashi, M., Ohashi, N., \& Miyama, S.M. 1993, \apj, 418, L71
\reference{herb} Herbig, G.H., \& Jones, B.F. 1983, \aj, 88, 1040
\reference{hold} Holdaway, M.A., \& Owens, F.N. 1995, NRAO. Millimeter Array Memo 126. 
\reference{jenn} Jennings, R.E., Cameron, D.H.M., Cudlip, W., \& Hirst, C.J. 1987, \mnras, 226, 461
\reference{j94} Jensen, E.L.N., Mathieu, R.D., \& Fuller, G.A. 1994, \apj, 429, L29
\reference{j96} Jensen, E.L.N., Mathieu, R.D., \& Fuller, G.A. 1996, \apj, 458, 312
\reference{jone} Jones, B.F., \& Cohen, M. 1986, \apj, 311, L23
\reference{ken} Kenyon, S.J,, \& Hartmann, L. 1987, \apj, 323, 714
\reference{kit} Kitamura, Y., Kawabe, R., \& Saito, M. 1996, \apj, 456, L137
\reference{k99} Koerner, D.W., \& Sargent, A.I., 1999, in preparation
\reference{k93} Koerner, D.W., Sargent, A.I., \& Beckwith, S.V.W. 1993, Icarus, 106, 2
\reference{l96} Lada, C.J., Alves, J., \& Lada, E.A. 1996, \aj, 111, 1964
\reference{l84} Lada, C.J., \& Wilking B.A. 1984, \apj, 287, 610
\reference{lar69} Larson, R.B. 1969, \mnras, 145, 271
\reference{lar78} Larson, R.B. 1969, \mnras, 184, 69
\reference{lar95} Larson, R.B. 1995, \mnras, 272, 213
\reference{lan} Langer, W.D., Castets, A., \& Lefloch, B. 1996, \apj, 471, L111
\reference{lay95} Lay, O.P., Carlstrom, J.E., \& Hills, R.E. 1995, \apj, 452, L73
\reference{lay97} Lay, O.P., Carlstrom, J.E., \& Hills, R.E. 1997, \apj, 489, 917
\reference{lay94} Lay, O.P., Carlstrom, J.E., Hills, R.E., \& Philips, T.G. 1994, \apj, 434, L75
\reference{la} Lavalley, C., Cabrit, S., Dougados, C., Ferruit, P., \& Bacon, R. 1997, \aap, 327, 671
\reference{le91} Leinert, Ch., Haas, M., Richichi, A., Zinnecker, H., \& Mundt, R. 1991, \aap, 250, 407
\reference{le93} Leinert, Ch., Zinnecker, H., Weitzel, N., Christou, J., Ridgway, S.T., Jameson, R., Haas, M., \& Lenzen, R., 1993, \aap, 278, 129
\reference{lis} Liseau, R., Sandell, G., \& Knee, L.B.G. 1988, \aap, 192, 153
\reference{me} Looney, L.W., Mundy, L.G., \& Welch, W.J. 1997, 484, L157
\reference{mi97} Mitchell, G.F., Sargent, A.I., \& Mannings, V. 1997, \apj, 483, L127
\reference{mi94} Mitchell, G.F., Hasegawa, T.I., Dent, W.R.F., \& Matthews, H.E. 1994, \apj, 436, L177
\reference{mon} Monaghan, J.J., \& Lattanzio, J.C. 1986, \aap, 158, 207
\reference{mous} Mouschovias, T.Ch. 1991, \apj, 373, 169
\reference{mun90} Mundt, R., Ray, T.P., Raga, A.C., \& Solf, J. 1990, \aap, 232, 37
\reference{mun87} Mundt, R., Brugel, E.W., \& B\"{u}hrke, T. 1987, \apj, 319, 275
\reference{mun83} Mundt, R., \& Fried, J.W. 1983, \apj, 274, L83
\reference{mund99} Mundy, L.G., Looney, \& Welch, W.J. 1999, in preparation
\reference{mund96} Mundy, L.G., Looney, L.W., Erickson, W., Grossman, A.W., Welch, W.J., Forster, J.R., Wright, M.C.H., Plambeck, R.L., Lugten, J., \& Thornton, D.D. 1996, \apj, 464, L169
\reference{mund93} Mundy, L.G., McMullin, J.P., Grossman, A.W., \& Sandell, G. 1993, Icarus, 106
\reference{mund92} Mundy, L.G., Wootten, A., Wilking, B.A., Blake, G.A., \& Sargent, A.I. 1992, \apj, 385, 306
\reference{mund86} Mundy, L.G., Wilking, B.A., \& Myers, S.T. 1986, \apj, 311, L75
\reference{ohas} Ohashi, N., Kawabe, R., Hayashi, M., \& Ishiguro, M. 1991, \aj, 102, 2054
\reference{ost} Osterloh, M., \& Beckwith, S.V.W. 1995, \apj, 439, 288
\reference{pat99} Patience, J.,  Ghez, A.M., Reid, I.N., \& Matthews, K. 1999, in preparation
\reference{pat98} Patience, J., Ghez, A.M., Reid, I.N., Weinberger, A.J., \& Matthews, K. 1998, \aj, 115, 1972
\reference{pen69} Penston, M.V. 1969, \mnras, 144, 425
\reference{poll} Pollack, J.B., Hollenbach, D., Beckwith, S.V.W., Simonelli, D.P., Roush, T., \& Fong, W. 1994, \apj, 421, 615
\reference{pring} Pringle, J.E. 1989, \mnras, 239, 361
\reference{pud} Pudritz, R.E., Wilson, C.D., Carlstrom, J.E., Lay, O.P., Hills, R.E., \& Ward-Thompson, D. 1996, \apj, 470, L123
\reference{rei} Reipurth, B., \& Zinnecker, H. 1993, \aap, 278, 81
\reference{rod97} Rodr\'{i}guez, L.F., Anglada, G., \& Curiel, S. 1997, \apj, 480, L125
\reference{rod95} Rodr\'{i}guez, L.F., Anglada, G., \& Raga, A. 1995, \apj, 454, L149
\reference{rod94} Rodr\'{i}guez, L.F., Cant\'{o}, J., Torrelles, J.M., G\'{o}mez, J.F., Anglada, G., i\& Ho, P.T.P 1994, \apj, 427, L29
\reference{rod98} Rodr\'{i}guez, L.F., D'Alessio, P., Wilner, D.J., Ho P.T.P., Torrelles, J.M., Curiel, S., Gomez, Y., Lizano, S., Pedlar, A., Canto, J., \& Raga, A.C. 1998, \nat, 395, 355
\reference{sar} Sargent, A.I., \& Beckwith, S.V.W. 1991, \apj, 382, L31
\reference{san91} Sandell, G., Aspin, C., Duncan, W.D., Russell, A.P.G., \& Robson, I.E. 1991, \apj, 376, L17
\reference{san94} Sandell, G., Knee, L.B.G., Aspin, C., Robson, I.E., \& Russell, A.P.G. 1994, \aap, 285, L1
\reference{sim92} Simon, M., Chen, W.P., Howell, R.R., Benson, J.A., \& Slowik, D. 1992, \apj, 384, 212
\reference{sim92b} Simon, M., \& Guilloteau, S. 1992, \apj, 397, L47
\reference{shu77} Shu, F.H. 1977, \apj, 214, 488 
\reference{shu93} Shu, F.H., Najita, J., Galli, D., Ostriker, E., \& Lizano, S. 1993, Protostars and Planets III, ed. E.H. Levy \& J.I. Lunine, (Tucson, Univ. of Arizona Press), 3
\reference{shu90} Shu, F.H., Tremaine, S., Adams, F.C., \& Ruden, S.P 1990, \apj, 358, 495
\reference{snell} Snell, R.L., Loren, R.B., \& Plambeck, R.L. 1980, \apj, 239, L17
\reference{stah} Stahler, S.W., Korycansky, D.G., Brothers, M.J., \& Touma, J. 1994, \apj, 431, 341
\reference{sta97} Stapelfeldt, K.R., Burrows, C.J., Krist, J.E., \& the WFPC2 Science Team 1997, IAU Symposium 182, Herbig-Haro Flows and the Birth of Low Mass Stars, ed. B. Reipurth \& C. Bertout, (Dordrecht: Kluwer), 355
\reference{sta95} Stapelfeldt, K.R., et al. 1995, \apj, 449, 888
\reference{stog} Stognienko, R., Henning, T., \& Ossenkopf, V. 1995, \aap, 296, 797
\reference{str76} Strom, S.E., Vrba, F.J., \& Strom, K.M., 1976, \aj, 81, 314
\reference{str74} Strom, S.E., Grasdalen, G.L., \& Strom, K.M. 1974, \apj, 191, 111
\reference{ter97} Terebey, S., \& Padgett, D.L. 1997 , IAU Symposium 182, Herbig-Haro Flows and the Birth of Low Mass Stars, ed. B. Reipurth \& C. Bertout, (Dordrecht: Kluwer), 507
\reference{ter93} Terebey, S., Chandler, C.J., \& Andr\'{e}, P. 1993, \apj, 414, 759
\reference{walk93} Walker, C.K., Carlstrom, J.E., Bieging, J.H., Lada, C.J., \& Young, E.T. 1990, \apj, 364, 173
\reference{walk} Walker, C.K., Lada, C.J., Young, E.T., Maloney, P.R., \& Wilking, B.A. 1986, \apj, 309, L47
\reference{war} Warin, S., Castets, A., Langer, W.D., Wilson, R.W., \& Pagani, L. 1996, \aap, 306, 935
\reference{wein} Weintraub, D.A., Kastner, J/H., \& Whitney, B.A. 1995, \apj, 452, L141
\reference{welch} Welch, W.J. et al. 1996, \pasp, 108, 93
\reference{welch99} Welch, W.J., Looney, L.W., \& Mundy, L.G. 1999, in preparation
\reference{whit} Whittet, D.C.B. 1974, \mnras, 168, 371
\reference{wil96} Wilner, D.J., Ho, P.T.P., \& Rodr\'{i}guez, L.F. 1996, \apj, 470, L117
\reference{wil99} Wilner, D.J., Rodr\'{i}guez, L.F., \& Ho, P.T.P. 1999, in preparation
\reference{woot} Wootten, A. 1989, \apj, 337, 858
\reference{yu} Yu, T., \& Chernin, L.M. 1997, \apj, 479, L63
\end{references}
\end{document}